\documentclass[12pt]{amsart}\usepackage{a4wide,enumerate,color,graphicx}
\usepackage{layouts}
\usepackage{amsmath}
\usepackage{scrextend} 
\usepackage{amssymb}
\usepackage{esint}
\usepackage{amsfonts}
\usepackage{amsthm}
\usepackage{enumerate}
\usepackage{mathrsfs}
\usepackage{caption}
\usepackage{url,hyperref}
\usepackage[scaled=0.92]{helvet}  

\DeclareMathOperator{\divv }{div}

\DeclareMathAlphabet{\mathdutchcal}{U}{dutchcal}{m}{n}

\newcommand{\calc}{\ensuremath\mathdutchcal{c}}

\newcommand{\calf}{\ensuremath\mathdutchcal{f}}
\newcommand{\calg}{\ensuremath\mathdutchcal{g}}

\newcommand{\calx}{\ensuremath\mathdutchcal{x}}
\newcommand{\calv}{\ensuremath\mathdutchcal{v}}

\newcommand{\calk}{\ensuremath\mathdutchcal{k}}
\allowdisplaybreaks

\begin{document}

\title[Compressibility and volume variations]{Compressibility and volume variations due to composition in multicomponent fluids}


\author[P.-E.~Druet]{Pierre-Etienne Druet}

\date{\today}

\maketitle

\begin{abstract}
For single components fluids, vanishing isothermal compressibility implies that the mass density is constant, but the same conclusion is unknown for multicomponent fluids. Here the volume remains affected by changes of the composition. In the present paper we discuss an apparently natural way to conceptualise, based on derivatives of the Gibbs function $g$, this \textquotesingle volume change due to composition\textquotesingle{} as a compression. In this way the phenomenon becomes quantitatively comparable to the usual coefficients of compressibility. This is a first step to investigate the range of validity of the constant density approximation of multicomponent fluids. As an illustration, three different aqueous solutions are discussed. 
\end{abstract}



\section{Introduction}

According to the thermal equation of state, the density $\varrho$ of a single component fluid (the specific volume $\calv = 1/\varrho$) is a function of temperature $T$ and pressure $p$, and the isothermal compressibility is defined via
\begin{align*}
\beta_T = - \frac{1}{\calv(T,p)} \,  \partial_p \calv(T,p) \, .
\end{align*}
If $\beta_T = 0$, that is \textquotedblleft compressibility equals to zero \textquotedblright{}, then the fluid is {\it in}compressible. Its density also does not depend on temperature -- even if this last conclusion should not be absolutely valid, \cite{gouin2012muller,zbMATH06469617,zbMATH07824760}. Hence, the density of a fluid with zero compressibility is constant, at least in a certain range of thermodynamic conditions.

In a multicomponent fluid, the volume varies with temperature and pressure but also with the composition. In this paper, the composition shall be expressed by the mass fractions of the components, denoted by $y_1, \ldots, y_N$ if we consider a mixture with $N \geq 2$ constituents ${\rm A}_1, \ldots, {\rm A}_N$. For instance, a solution of water ${\rm A}_1$ and ethanol ${\rm A}_2$ under standard temperature and pressure conditions possesses at $y_1 =1$ a density of $997 \, {\rm kg}/{\rm m}^3$ and at $y_1 = 0$ of $798 \, {\rm kg}/{\rm m}^3$. Of course, we would at first not think of this phenomenon as a compression.

Nonetheless, in the multicomponent case, the isothermal compressibility is likewise defined as
\begin{align}\label{DEF00}
	\beta_T = \beta_{(T,y)} = - \frac{1}{\calv(T,p, \, y_1, \ldots, y_N)} \,  \partial_p \calv(T,p, \, y_1, \ldots, y_N) \, .
\end{align}
Here however, $\beta_T = 0$ does not lead to the conclusion that the volume or the mass density is constant. In the paper \cite{zbMATH07824760} we show that $\beta_T = 0$ leads, even under maximal assumptions, to a representation $\calv(T,p,y_1, \ldots,y_N) = \sum_{i = 1}^N \bar{\calv}_i \, y_i$ with constants $\bar{\calv}_1, \ldots, \bar{\calv}_N$. This means that the volume is independent of temperature and pressure, but continues to depend on the composition. This conclusion looks natural, since the measure defined in \eqref{DEF00} takes only pressure variations into account.

In the present paper, we discuss the possibility to measure the relative importance of the true compression effects, measured by $\beta_T$ (or, more exactly, $\beta_{(T,y)}$) and the volume effects resulting from composition. In order to quantify the latter we consider the function (see below formula \eqref{FUND1})
\begin{align}\label{DEF01}
\calk = \calk(T,p, {\bf y}) :=	\partial_pg(T,p,{\bf y}) \,   \big(\mathbf{D^2_{{\bf y},{\bf y}}} g(T,p,{\bf y})\big)^{-1} \Big(\frac{1}{\calv} \, \partial_{\bf y} \calv\Big) \cdot \Big(\frac{1}{\calv} \, \partial_{ \bf y} \calv\Big)  \, ,
\end{align}
in which $g$ is the Gibbs function, and derivatives with respect to ${\bf y}$ are taken tangentially to the hyperplane $\sum_{i=1}^N y_i = 1$. We will show that this coefficient occurs naturally as a compressibility if instead of fixing the composition in \eqref{DEF00} we fix the chemical potentials. In fact, we show that
\begin{align*}
	\beta_{(T,y)} + \calk \stackrel{\rm !}{=} -\frac{1}{\calv(T,p,\mu_1-\mu_N, \ldots,\mu_{n-1}-\mu_N)} \, \partial_p\calv(T,p,\mu_1-\mu_N, \ldots,\mu_{n-1}-\mu_N)
\end{align*}
with the vector $\boldsymbol{ \mu} = (\mu_1, \ldots,\mu_N)$ of chemical potentials. With the help of its projection $\boldsymbol{\mu}^t$ onto the tangent hyperplane of $\sum_{i=1}^N y_i = 1$, the overall compressibility coefficient of the fluid mixture is
\begin{align}\label{betatrinsic}
	\beta_{(T, \, \mu^t)} = -\frac{1}{\calv(T,p,\boldsymbol{\mu}^t)} \, \partial_p\calv(T,p,\boldsymbol{\mu}^t) = \beta_{(T,y)} + \calk
\end{align}
We shall moreover illustrate the theory with three examples  of aqueous solutions: Water + salt, water + sucrose and water + ethanol. In the two first cases, the ratio of $\calk/\beta_{(T,y)}$ can become large for a dense mixture while, for the third case, the maximal ratio remains slightly above $1$ over the whole range of compositions. 

Our conclusion is that we should attempt to distinguish multicomponent incompressible fluids from multicomponent constant density fluids. An quantitative characterisation of the difference could be based on the ratio $\calk/\beta_{(T,y)}$. Obviously, this will have to be studied further using wider data sets and more performant computational toolboxes where full Gibbs functions are available.

Our plan is as follows. In the section \ref{DEFS} we introduce the basic thermodynamic setting and recall some definitions. Then in the section \ref{COMPUTATIONRULES} we derive the computation rules for the derivatives of functions depending on density/volume and chemical potentials. We show how the definition \eqref{DEF01} arises as a pressure derivative of density, hence a compressibility. The section \ref{EXAMPES} is devoted to the three illustrations. Finally, in an appendix, we have collected some auxiliary materials and complements.

\section{The thermodynamic setting}\label{DEFS}

\subsection{A few definitions}
We shall use the thermodynamic setting of non-equilibrium thermodynamics to describe the mixture. We refer to the work of W.~Dreyer, for instance to the Appendix of \cite{zbMATH07824760} or to \cite{zbMATH06469617} about reactive fluid mixtures.

{\bf Variables.} The thermodynamic conditions are locally expressed by $N+1$ variables
\begin{align}\label{basicvariables}
	T \,\, \text{-- absolute temperature,} \quad {\boldsymbol \rho} = (\rho_1, \ldots,\rho_N) \,\, \text{ -- partial mass densities.}
\end{align}
The partial mass density $\rho_i$ is the amount of mass of ${\rm A}_i$ available per unit volume of the mixture. The mass density of the fluid, is defined as
\begin{align}\label{rhotot}
	\varrho = \sum_{i=1}^N \rho_i \, \, \text{-- total mass density,} \quad y_i = \frac{\rho_i}{\varrho} \,\, \text{-- $i^{\rm th}$ mass fractions.}
\end{align}
Beside the main variables $T$ and $\rho_1, \ldots,\rho_N$ introduced in \eqref{basicvariables}, different usual sets of thermodynamic variables are encountered. 
The partial mass densities and the internal energy density, denoted by $\varrho u$, are sometimes called \emph{conservative variables} (\cite{zbMATH06040068}, Ch.\ 8 and 9)
\begin{align}\label{variables1}
	(\rho_1,\ldots,\rho_N, \,\varrho u) \,\, \text{ -- conservative variables.}
\end{align}
Further, with the molar masses (or with the molecular masses) $M_1, \ldots, M_N > 0$ of the constituents, assumed to be constants throughout this investigation, we introduce 
\begin{align}\label{molardens}
	n_i := \frac{\rho_i}{M_i} \, ,& \qquad n = \sum_{i= 1}^N n_i \,, \qquad
	x_i = \frac{n_i}{n}  \, ,
\end{align}
and we call $n_i = $ the partial mole density (the number density), $n$ the total mole density (the number density) and $x_i =$ the mole fraction (the number fraction). Another set of variables uses the temperature, the pressure and the mole or number fractions. These are the variables of the specific Gibbs energy $g$ which are typically the quantities controlled in experiments:
\begin{align}\label{variables2}
	(T, \, p, \, x_1, \ldots,x_N) \, .
\end{align}
The set \eqref{variables2} occurs in connection with the discussion of ideal mixtures \eqref{muiideal} and in the illustration of Section \ref{EXAMPES}. Owing to the relationships
\begin{align*}
y_i = \frac{M_i \, x_i}{\sum_{j=1}^N M_j \, x_j}, \, \quad x_i = \frac{y_i}{M_i \, \sum_{j=1}^N(y_j/ M_j)} \, ,
\end{align*}
a set of variables completely equivalent ot \eqref{variables2} is given as
\begin{align}\label{variables3}
	(T, \, p, \, y_1, \ldots,y_N) \, .
\end{align}

{\bf Thermodynamic stability and the consequences.} In the \emph{stable fluid phase}, the sets of state variables \eqref{basicvariables}, \eqref{variables1}, \eqref{variables2} and \eqref{variables3} are all equivalent, in the sense that there are smooth bijections transforming these vectors into one another (cp.\ among others \cite{zbMATH07824760}, \cite{zbMATH06049914}). We let $\varrho s$ be the bulk density of the mixture entropy possesses of the special form
\begin{align}\label{ENTROPE}
\varrho s = - h(\rho_1, \ldots, \rho_N, \, \varrho u) \, ,
\end{align}
where $s$ is the specific entropy and $h: \, \mathcal{D} \subseteq \mathbb{R}^N_+ \times \mathbb{R} \rightarrow \mathbb{R}$ with $\mathbb{R}^N_+ := ]0, \, +\infty[^N$ the (problem dependent) given constitutive function, assumed strictly convex and sufficiently smooth in its domain $\mathcal{D}$. It is usual to call $h$ the mathematical entropy function -- which is convex. 

The concavity-postulate is usually derived from the requirement of increasing entropy in all thermodynamic processes, as for instance in the first chapter of the book \cite{mueller}. In \cite{zbMATH06469617}, Section 4, number 7., another motivation of the concavity postulate is given, starting from a stability principle for an homogeneous system with variable volume. 
As W.~Dreyer used to underline, it is wrong to infer that a convex $h$ alone guarantees that the entropy is a stability functional for a thermodynamic process. This is mainly true only for insulated systems. 
The following definitions relate the (mass-based) chemical potentials $\mu_1, \ldots,\mu_N$ and the internal energy density $\varrho u$ to the state variables
\begin{align}
\label{TEMPCHEMPOT} -\frac{1}{T} & = \partial_{\varrho u} h( \rho_1, \ldots, \rho_N, \, \varrho u)  \quad \text{ and } \quad  \frac{\mu_i}{T} =  \partial_{\rho_i}h(\rho_1, \ldots, \rho_N, \, \varrho u) \, \, \text{ for } i = 1,\ldots,N \, .
\end{align}
Further, the thermodynamic pressure $p$ obeys the \emph{Gibbs-Duhem equation}
\begin{align}\label{GIBBSDUHEMEULER}
p = -\varrho u + T \, \varrho s  + \sum_{i=1}^N \rho_i \, \mu_i  \, .
\end{align}
The Helmholtz free energy is defined via $\varrho \psi := \varrho u - T \, \varrho s$. Its constitutive function has the form
\begin{align}\label{thermo8a}
\varrho \psi = f(T, \, \rho_1 ,  \ldots, \rho_N ) \, ,
\end{align}
and then the (mass--based) chemical potentials are given by\footnote{In \cite{zbMATH06040068}, see Sec.\ 6.2.3, the same object is called {\it species Gibbs function} denoted by $\mathcal{G}_i$, while the notion of {\it chemical potential} with symbol $\mu_i$ is reserved for the quantity $\mathcal{G}_i/(RT)$.}
\begin{align}\label{thermo8b}
\mu_i = \partial_{\rho_i} f \quad \text{ for } i  = 1, 2, \ldots , N \, .
\end{align}
Other basic quantities can be calculated from $f$, via
\begin{gather}\label{thermo8c}
p = -f +
\sum_{i=1}^N \rho_i \, \mu_i  \,  , \quad \varrho s = - \partial_T f \, \\
\varrho u = - T^2 \, \partial_T \frac{f}{T} =: \epsilon(T, \, \rho_1,\ldots,\rho_N) \, ,
\end{gather} 
If the constitutive functions $f$ and $h$ are twice differentiable, the following relationships are valid
\begin{align}& \label{relHessiansrho}
\mathbf{D^{2}_{\boldsymbol{\rho},\boldsymbol{\rho}}}f(T, \, \boldsymbol{\rho}) = T \,  \Big(\mathbf{D^{2}_{\boldsymbol{\rho},\boldsymbol{\rho}}}h(\boldsymbol{\rho}, \, \varrho u) - \frac{\mathbf{D^2}_{\boldsymbol{\rho},\varrho u} h(\boldsymbol{\rho}, \, \varrho u) \otimes\mathbf{D^2}_{\boldsymbol{\rho},\varrho u} h(\boldsymbol{\rho}, \, \varrho u)}{\partial^2_{\varrho u} h(\boldsymbol{\rho}, \, \varrho u)}\Big)\\
& \label{relHessianT}\partial^2_{T} f(T, \, \boldsymbol{\rho}) = - \frac{\partial_T \epsilon(T, \, \boldsymbol{\rho})}{T}  = - \frac{1}{T^3 \, \partial^2_{\varrho u} h(\boldsymbol{\rho}, \, \varrho u)} \, .
\end{align}
In the main variables \eqref{basicvariables}, stability is hence expressed by the two positivity conditions
\begin{align*}
\{\partial^2_{\rho_i,\rho_j} f\} > 0 \quad \text{ and } \quad \partial_T\epsilon > 0\, ,
\end{align*}
where $>0$ applied to a matrix means positive definite. Another important function is the specific Gibbs free enthalpy $g = \psi + p/\varrho $ using the variables \eqref{variables3} hence
\begin{align}\label{Gibbs}
	g = g(T, \, p, \, y_1, \ldots, y_N) \, ,
\end{align} 
where the derivatives satisfy
\begin{align}\label{Gibbsderiv1}
\partial_T g =-s, \quad \partial_{p} g = \frac{1}{\varrho} \, .
\end{align}	
The compositional, $y-$derivative is special, because it makes sense only along the hypersurface $\sum_{i=1}^N y_i = 1$. Hence, it is a tangential derivative, which we express with the symbol $\partial^t_{\bf y}$, and get
\begin{align}\label{gibbsyderivgen}
	\partial^t_{\bf y} g= \boldsymbol{\mu}^t \, .
\end{align}
Using particular parametrisations of the hypersurface, for instance $y_N = 1-\sum_{j=1}^{N-1} y_j$, we get special representation of the tangent vector $\boldsymbol{ \mu}^t$:
 \begin{align}\label{gibbsyderivpart}
 	\partial^t_{y_i}g = \mu_i-\mu_N \quad \text{ for } \quad i = 1,\ldots,N-1 \, .
 \end{align}
 However, in the general meaning expressed by \eqref{gibbsyderivgen}, $\boldsymbol{ \mu}^t$ is the $N-$dimensional projection of the vector of chemical potentials onto the orthogonal complement of the one vector $\mathbf{1} = (1, \ldots,1)^{\sf T}$. For the Gibbs function the stability conditions then read
 \begin{align*}
 	\{\partial^2_{(T,p)} g\} < 0 \quad  \text{ and } \quad \{\partial^{2}_{y_i,y_j} g\}^t > 0 \, .
 \end{align*}


{\bf Ideal mixture.} Following M\"uller, for instance in eq.\ (7.22) of \cite{muller2013grundzuge}, the concept of an ideal mixture refers to a concrete form of the chemical potentials
\begin{align}\label{muiideal}
	\mu_i = \mu_i(T, \, p, \, x_i) = g_i(T, \, p) + \frac{R \, T}{M_i} \, \ln x_i \, .
\end{align}
Here $R$ is the gas constant. In \eqref{muiideal}, $g_i(T, \, p)$ denotes the Gibbs free enthalpy of the constituent ${\rm A}_i$. In particular, with $\hat{\rho}_i(T, \, p)$ being the mass density of the constituent ${\rm A}_i$ as a function of temperature and pressure, we have the identity $\partial_p g_i(T, \, p) = 1/\hat{\rho}_i(T, \, p) =: \calv_i(T,\, p)$ with the specific volume of ${\rm A}_i$. How to construct a full thermodynamic model (particular constitutive models) respecting \eqref{muiideal} was shown in \cite{zbMATH07554819}, Section 3.

From the Gibbs--Duhem--Euler equation \eqref{GIBBSDUHEMEULER} and from \eqref{muiideal} we can derive the identity
\begin{align}\label{VOLADD}
\sum_{i=1}^N \rho_i \, \partial_pg_i(T,\, p) = 1 \, ,
\end{align}
which we call the volume additivity of ideal mixtures. It implicitly characterises the pressure as a function of $(T, \, \rho)$, hence we can call the latter relation the equation of state of ideal mixtures.

{\bf Compressibility and heat capacity.} The isothermal and isocompositional comppressibility is defined as $\beta_{(T,y)} = \partial_p \varrho/\varrho$, where the variables \eqref{variables3} are considered. Hence the $p-$derivative is built while the temperature $T$ and the composition vector ${\bf y}$ are fixed. Using the Gibbs function and in particular \eqref{Gibbsderiv1}$_2$, we have
\begin{align}\label{compressibilityprimaer}
\beta_{(T,y)} = - \frac{\partial^2_p g(T,p, \, {\bf y})}{\partial_p g(T,p, \, {\bf y})} > 0 \quad \text{ -- isothermal, isocompositional compressibility, }
\end{align}	
where the inequality follows from thermodynamic stability. Using the Helmholtz function $f$, we have equivalently
\begin{align}\label{isocomp}
& \beta_{(T, \, y)} = \frac{1}{\mathbf{D^2_{{\boldsymbol{\rho}},{\boldsymbol{\rho}}}}f(T,{\boldsymbol{\rho}})  {\boldsymbol{\rho}} \cdot {\boldsymbol{\rho}}} \, , 
\end{align}
while \eqref{relHessiansrho} can be used to express $\beta_{(T, \, y)}$ in terms of the entropy potential.

Another important function is the isothermal and isocompositional heat capacity at constant volume, which is primarily defined as $\calc_{(T,y)} = \partial_T \varrho u(T, \, {\bf \rho})$, hence
\begin{align}
& \label{heatcapacity} \calc_{(\upsilon,y)} :=  \frac{1}{\varrho} \, \partial_T \epsilon(T, \, {\boldsymbol{\rho}}) > 0 \quad \text{ -- heat capacity at constant volume and composition.}
\end{align}
For an ideal mixture, these quantities can be computed from the data $g_1,\ldots,g_N$ of the species, and adopt the following form (see \cite{zbMATH07554819}, Lemma 3.1)
\begin{align}\label{isocompideal}
	& \beta_{(T, \, y)} = - \sum_{i=1}^N \rho_i \, \partial_p^2 g_i(T, \, p) \, ,\\
	& \calc_{(\upsilon,\, y)} = - \frac{T}{\varrho} \, \left(\sum_{i=1}^N\partial^2_{T} g_i(T, \, p) \, \rho_i -\frac{\Big(\sum_{i=1}^N \partial^2_{T,p} g_i(T, \, p) \, \rho_i\Big)^2}{\sum_{i=1}^N \partial^2_{p} g_i(T, \, p) \, \rho_i} \right) \, .
\end{align}
With the isothermal compressibility of the species given as
\begin{align*}
	\beta^i_{T}(T,p) := -\hat{\rho}_i(T,p) \, \partial^2_pg_i(T, \, p) \, ,
\end{align*}
and with the volume fractions $\phi_i = \rho_i/\hat{\rho}_i(T,p)$ which sum up to one owing to \eqref{VOLADD}, we can write \eqref{isocompideal}$_1$ as
\begin{align*}
\beta_{(T, \, y)} = \sum_{i=1}^N \phi_i \, \beta^i_{T}(T,\, p) \, .
\end{align*}
The relationship between the heat capacity $\calc_{(\upsilon,y)}$ and the heat capacities of the species at constant volume is more complex, see Lemma 3.1 in \cite{zbMATH07554819}.

\subsection{Entropic (normal form) variables}

For I.~Müller, the chemical potentials (together with the temperature) are --more fundamentally than the partial densities $\rho$, the pressure or the composition vector $\bf y$ -- the continuous quantities expressing the equilibrium state of the mixture, which one can measure. And thus \textquotedblleft (...) we have to get acquainted with the chemical potential, even if that is not pleasant\textquotedblright, see section 7.1.2 in \cite{muller2013grundzuge}.

The Theory of Irreversible Processes postulates that the mass and heat fluxes in a nonequilibrium multicomponent systems are proportional to gradients of chemical potentials and temperature, hence these are the driving forces thermodynamic systems in non equilibrium. For a recent introduction to multicomponent diffusion, let us refer to \cite{zbMATH06469617} and \cite{zbMATH07653138} where the reader can find many references to the primary literature. In the general theory of diffusion there is the subtle point that not $\mu_1,\ldots,\mu_N$ drive the mass and heat diffusion, but rather $\boldsymbol{ \mu}^t$ (see \eqref{gibbsyderivgen}, \eqref{gibbsyderivpart}).
Since the mass fractions are subject to the condition $\sum_{i=1}^N y_i = 1$, the vector $\boldsymbol{\mu}^t$ has zero average and possesses only $N-1$ independent components. Particular representations can be obtained from bases of the hyperplane $\sum_{i=1}^N y_i = 1$, for instance $\boldsymbol{\eta}^k = {\bf e}^k-{\bf e}^N$ for $k = 1,\ldots,N-1$, with the Euclidian basis vectors. In this case
\begin{align}\label{chempot3}
\boldsymbol{\eta}^i \cdot \boldsymbol{\mu}^t = \mu_i - \mu_N \quad \text{ for } \quad i = 1, \ldots, N-1 \, .
\end{align} 
With the paper \cite{C3CP44390F} we see that one can speak of $\boldsymbol{\mu}^t$ as ''relative chemical potentials''.

In the context of mathematical treatment for the partial differential equations of multicomponent fluid dynamics \cite{zbMATH07554819}, it has proven convenient to separate the mass density $\varrho$ subject to the continuity equation (hyperbolic conservation law) and $N-1$ variables contributing to entropic dissipation (parabolic system):
\begin{alignat}{2}
	\label{parabvar} & \left. \begin{matrix}
		\frac{\mu_1 - \mu_N}{T}, \ldots,  \frac{\mu_{N-1} - \mu_N}{T} & \text{ (relative) chemical potentials}\\[0.5ex]
		-\frac{1}{T}  & \text{ reciprocal of temperature}
	\end{matrix} \right\} \quad & & \text{ the parabolic variables,}\\[0.1ex]
	\label{hyperbvar} & \varrho \, \, \text{ total mass density}  & & \text{the hyperbolic variable.}
\end{alignat}
Up to the choice of $-1/T$ instead of $T$, this set of variables belongs to the {\it intermediate normal form} of multicomponent fluid dynamics: See \cite{zbMATH06040068}, Section 8.7 for more details.

{\bf The transformation rules.} Following \cite{zbMATH07554819}, we shall in fact introduce the change of variables in a slightly more general way than \eqref{parabvar}. This turns especially useful in the analysis of incompressible models, see \cite{zbMATH07330765,zbMATH07698266} and \cite{zbMATH07452670}. First we define the $(N+1)-$vector of extended state variables
\begin{align}\label{Defw}
	w_i := \rho_i \quad \text{ for } i =1,\ldots,N \,, \quad \quad  w_{N+1} := \varrho u \,,
\end{align}
to denote the conservative variables \eqref{variables1}, the ones occurring in the entropy functional. Exploiting the relations \eqref{TEMPCHEMPOT}, the combinations
\begin{align}\label{Defwprime}
	w^*_i := \frac{\mu_{i}}{T} \quad \text{ for } i = 1, \ldots, N \, , \quad \quad w^*_{N+1} := - \frac{1}{T}\, ,
\end{align}
are the dual variables, also called the ''entropic variables''.
%
Suppose that the entropy function of \eqref{ENTROPE} is strict convex in its domain. Then, with the help of the conjugate convex function $h^*$ to $h$, we can invert the relations 
\begin{align*}
	\frac{\mu_i}{T} = \partial_{\rho_i}h(\boldsymbol{\rho}, \, \varrho u) \quad \text{ for } i=1,\ldots,N \, , \quad \quad 
	-\frac{1}{T} = \partial_{\varrho u} h(\boldsymbol{\rho}, \, \varrho u) \, ,
\end{align*}
which more compactly now read as 
\begin{align}\label{wtowstar}
	\mathbf{ w^*} = \boldsymbol{\nabla}_{\mathbf{w}} h (\mathbf{w}) \, . 
\end{align}
In order to guarantee the possibility to switch between conservative and entropic variables, the notion of a function of Legendre-type is essential. We assume that $h$ is such a function of Legendre--type on $\mathcal{D} \subseteq \mathbb{R}^N_+ \times \mathbb{R}$ open, convex. We define $\mathcal{D}^*$ to be the image of $\boldsymbol{\nabla}_{\mathbf{w}} h$ on $\mathcal{D}$. If $\mathcal{D}^*$ is open and convex, then the conjugate $h^*(w^*) := \sup_{w \in \mathcal{D}} \{w^* \cdot w - h(w)\}$ is a function of Legendre--type on $\mathcal{D}^*$. The gradients $\boldsymbol{\nabla}_{\mathbf{w}} h$ on $\mathcal{D}$ and $\boldsymbol{\nabla}_{\mathbf{w^*}} h^*$ on $\mathcal{D}^*$ are inverse to each other, see for instance the lemma 2.3 of \cite{zbMATH07554819}. 

By means of a fixed linear transformation, we next separate the parabolic and hyperbolic variables. To this aim, we choose new axes $\boldsymbol{\xi}^1,\ldots,\boldsymbol{\xi}^{N}, \, \boldsymbol{\xi}^{N+1}$ of $\mathbb{R}^{N+1}$ in the following way:
\begin{align}\label{labase}
	\left\{
	\begin{matrix} 
		\boldsymbol{ \xi}^N :=  \mathbf{e}^{N+1} = (0, \ldots, \, 0, \, 1) \quad \text{ and } \quad \boldsymbol{\xi}^{N+1} := \boldsymbol{\bar{e}} := (1, \ldots, \, 1, \, 0) \, ,  \\[0.2cm]
		\boldsymbol{\xi}^1, \ldots, \boldsymbol{\xi}^{N-1} \in \Big\{{\bf \calx} \in \mathbb{R}^{N+1} \, : \, \calx_{N+1} = 0, \,  \sum_{i=1}^N \calx_i = 0\Big\} = \{\boldsymbol{\xi}^{N}, \,  \boldsymbol{\xi}^{N+1}\}^{\perp} \, .
	\end{matrix} \right.
\end{align}
As seen, typical is the choice $\boldsymbol{\xi}^k = \mathbf{e}^k- \mathbf{e}^N$ for $k = 1,\ldots,N-1$.
We let $\boldsymbol{\eta}^1,\ldots,\boldsymbol{\eta}^{N+1} \in \mathbb{R}^{N+1}$ be the dual basis for $\boldsymbol{\xi}^1,\ldots,\boldsymbol{\xi}^{N+1}$. Then
\begin{align}\label{etaspecia}
	\boldsymbol{\eta}^N = \mathbf{e}^{N+1}, \quad \boldsymbol{\eta}^{N+1} = \frac{1}{N} \, \boldsymbol{\xi}^{N+1}, \quad \eta^k_{N+1} = 0 \, \text{ and } \, \sum_{i=1}^{N}\eta^k_i = 0 \, \text{ for } \, k = 1,\ldots,N-1 \, .  
\end{align}
For $\mathbf{w^*} = (\mu_1/T, \ldots,\mu_N/T, \, -1/T)$, we define the projections (see \eqref{chempot3})
\begin{align}\label{relativepot}
	q_{\ell} := \boldsymbol{\eta}^{\ell} \cdot \mathbf{w^*} := \sum_{i=1}^{N+1} \eta^{\ell}_i \, w^*_i\quad  \text{ for } \quad  \ell = 1,\ldots,N \, .
\end{align}
Due to the properties of the chosen basis, in particular to \eqref{etaspecia}, we have a relationship 
\begin{align}\label{TRAFO1}
	& \mathbf{w^*} = \sum_{\ell = 1}^N q_{\ell} \, \boldsymbol{\xi}^{\ell} + (\mathbf{w^*} \cdot \boldsymbol{\eta}^{N+1}) \, \boldsymbol{ \xi}^{N+1}  \quad 
	\text{ implying that } \quad  w_{N+1}^* = q_{N} \, .
\end{align}
Now, since the coordinate $w_{N+1}^*$ has the physical meaning of $-1/T$, the relevant domain for the new variable $q$ is the half-space
\begin{align*}
	\mathcal{H}^N_- := \Big\{ (q_1, \ldots, q_N) \in \mathbb{R}^N \, :\, q_{N} < 0 \Big\} = \mathbb{R}^{N-1} \times \mathbb{R}_{-}  \, . 
\end{align*}
Since $( \mu_1/T, \ldots, \mu_N/T, \ -1/T) = \boldsymbol{\nabla}_{\mathbf{w}} h(\boldsymbol{\rho},\,  \varrho u)$, use of the conjugate convex function yields
\begin{align*}
	(\rho_1, \, \ldots, \rho_N, \varrho u) = \boldsymbol{\nabla}_{\mathbf{ w^*}} h^*(\mu_1/T, \ldots, \mu_N/T, \ -1/T) \, .
\end{align*}
In order to isolate the hyperbolic component $\varrho$ (total mass density), we now express
\begin{align*}
	\varrho = \sum_{i=1}^N \rho_i &  = \sum_{i=1}^{N+1} \xi^{N+1}_i \, \partial_{w^*_i} h^*(w^*_1,\ldots,w^*_{N+1}) \nonumber\\
	& = \boldsymbol{ \xi}^{N+1} \cdot \boldsymbol{\nabla}_{\mathbf{ w^*}}h^*\Big(\sum_{\ell = 1}^{N} q_{\ell} \, \boldsymbol{ \xi}^{\ell} + (\mathbf{ w^*} \cdot \boldsymbol{\eta}^{N+1}) \, \boldsymbol{ \xi}^{N+1}\Big) \, .
\end{align*}
This is an algebraic equation of the form $F(\mathbf{w^*} \cdot \boldsymbol{\eta}^{N+1}, \, q_1, \ldots, q_{N}, \, \varrho) = 0$. We notice that
\begin{align*}
	\partial_{\mathbf{w^*} \cdot \boldsymbol{\eta}^{N+1}} F(\mathbf{w^*} \cdot \boldsymbol{\eta}^{N+1}, \, q_1, \ldots, q_{N}, \, \varrho) = \mathbf{D^{2}}h^*(\mathbf{ w^*}) \boldsymbol{ \xi}^{N+1} \cdot \boldsymbol{ \xi}^{N+1} > 0 \, ,
\end{align*}
due to the strict convexity of the conjugate function. It can be shown (see Lemma 2.4 in \cite{zbMATH07554819}) that the latter algebraic equation defines the component $\mathbf{w^*} \cdot \boldsymbol{\eta}^{N+1}$ implicitly as a differentiable function of $\varrho$ and $q_1, \ldots, q_{N}$. We call this function $\mathscr{M}$, satisfying by definition
\begin{align}\label{rhoiso}
	\varrho = \boldsymbol{ \xi}^{N+1} \cdot \boldsymbol{\nabla}_{\mathbf{ w^*}}h^*\Big(\sum_{\ell = 1}^{N} q_{\ell} \, \boldsymbol{\xi}^{\ell} + \mathscr{M}(\varrho, \, \mathbf{q}) \, \boldsymbol{\xi}^{N+1}\Big) \, .
\end{align}
We obtain the equivalent formulae
\begin{align}\label{MUAVERAGE}
\mathbf{ w^*} & = \sum_{\ell=1}^{N} q_{\ell} \, \boldsymbol{ \xi}^{\ell} + \mathscr{M}(\varrho, \, q_1,\ldots,q_{N}) \, \boldsymbol{ \xi}^{N+1} \, ,\\
	\label{RHONEW}\rho_i & = \partial_{w^*_i}h^*\Big( \sum_{\ell=1}^{N} q_{\ell} \, \boldsymbol{ \xi}^{\ell} + \mathscr{M}(\varrho, \, q_1,\ldots,q_{N}) \, \boldsymbol{\xi}^{N+1}\Big) =: \mathscr{R}_i(\varrho, \, \mathbf{q}) \quad \text{ for } \quad i = 1,\ldots,N \, ,\\
	\label{RHOUNEW} \varrho u & = \partial_{w^*_{N+1}} h^*\Big( \sum_{\ell=1}^{N} q_{\ell} \, \boldsymbol{ \xi}^{\ell} + \mathscr{M}(\varrho, \, q_1,\ldots,q_{N}) \, \boldsymbol{ \xi}^{N+1}\Big) =: \mathscr{U}(\varrho, \, \mathbf{q})\, ,
\end{align}
with $\varrho$ and $q_1,\ldots,q_{N}$ as the free variables. Since $p$ obeys \eqref{GIBBSDUHEMEULER}, we have $p = T \, \calg(\boldsymbol{\nabla}_{\mathbf{w}} h(\mathbf{w}))$, and here $\calg$ denotes the Legendre transform of $h$. Since $\calg$ coincides with $h^*$ for a function $h$ of Legendre--type, we find that
\begin{align}\label{ptohstar}
	p = T \, h^*(\mathbf{w^*})  = - h^*(\mathbf{ w^*})/w_{N+1}^* \, .
\end{align}
We combine the latter with \eqref{MUAVERAGE} to obtain that
\begin{align}\label{PNEW}
	p = \mathscr{P}(\varrho, \, \mathbf{q}) := - \frac{1}{q_N} \,   h^*\Big( \sum_{\ell=1}^{N} q_{\ell} \, \boldsymbol{ \xi}^{\ell} + \mathscr{M}(\varrho, \, q_1,\ldots,q_{N}) \, \boldsymbol{ \xi}^{N+1}\Big) \, . 
\end{align}
Not only the pressure, but all thermodynamic quantities can now be introduced as functions of the variables $\varrho, \, \mathbf{q}$. Indeed, considering a function $\calf = \calf(T, \, \rho_1, \ldots, \rho_N)$ of the main variables, we use that $T = -1/q_N$ and $\rho = \boldsymbol{\mathscr{R}}(\varrho, \, \mathbf{q})$ (see \eqref{RHONEW}), and we define
\begin{align}\label{changetoentropic}
	\calf(\varrho, \, \mathbf{q}) := \calf\Big(-\frac{1}{q_N}, \, \mathscr{R}_1(\varrho, \, \mathbf{q}), \ldots, \mathscr{R}_N(\varrho, \, \mathbf{q})\Big)
\end{align}
to obtain the equivalent representation in the entropic variables.

\section{Transformed thermodynamic functions and their derivatives}\label{COMPUTATIONRULES}

In \eqref{relativepot}, we introduced the variables $q_1, \ldots,q_N$ as linear combinations of the entropic variables $w^*_1, \ldots,w^*_{N+1}$.
We next want to compute some derivatives of the transformed coefficient functions:
The map $\mathscr{R}(\varrho, \, \mathbf{q})$ of \eqref{RHONEW}, the function $\mathscr{U}(\varrho, \, \mathbf{q})$ of \eqref{RHOUNEW} and $\mathscr{P}(\varrho, \, \mathbf{q})$ of \eqref{PNEW}.
Recall that the fundament of the change of variables is the equation \eqref{rhoiso}.

\subsection{General formulas}

Differentiation directly gives the following expressions for the 
derivatives of $\mathscr{M}$:
\begin{align}\begin{split}\label{scriptmderiv}
	& \partial_{\varrho} \mathscr{M}(\varrho, \, \mathbf{q}) = \frac{1}{\mathbf{D^{2}}h^*(\mathbf{ w^*}) \boldsymbol{ \xi}^{N+1} \cdot \boldsymbol{ \xi}^{N+1}} \, ,\\
	& \partial_{q_k} \mathscr{M}(\varrho, \, \mathbf{q}) = - \frac{\mathbf{D^2}h^*(\mathbf{ w^*}) \boldsymbol{ \xi}^{N+1} \cdot \boldsymbol{ \xi}^k}{\mathbf{D^{2}}h^*(\mathbf{ w^*}) \boldsymbol{ \xi}^{N+1} \cdot \boldsymbol{ \xi}^{N+1}} \quad \text{ for } \quad k = 1,\ldots,N-1 \, ,\\
	& \partial_{q_N} \mathscr{M}(\varrho, \, \mathbf{q}) = - \frac{\mathbf{D^2}h^*(\mathbf{ w^*}) \boldsymbol{ \xi}^{N+1} \cdot \mathbf{e}^{N+1}}{\mathbf{D^2}h^*(\mathbf{ w^*}) \boldsymbol{ \xi}^{N+1} \cdot \boldsymbol{ \xi}^{N+1}} \, ,
\end{split}
\end{align}
where we recall that $\boldsymbol{ \xi}^{N+1} = (1,\ldots,1, \, 0)^{\sf T}$ and $\mathbf{ w^*} = \mathbf{ w^*}(\varrho, \, \mathbf{q})$ is given by \eqref{MUAVERAGE}. Note that the expression $\mathbf{D^{2}}h^*(\mathbf{ w^*}) \boldsymbol{ \xi}^{N+1} \cdot \boldsymbol{\xi}^{N+1}$ is strictly positive owing to the strict convexity of $h^*$. Moreover, we notice that the $\varrho-$derivative and the $q_N := - 1/T$ derivative are intrinsic expressions, while the $q_k$ derivative for $k=1,\ldots,N-1$ depends on the choice of the basis. 

Systematic computations are performed in the Appendix, Section \ref{derivatives}. Here we retain two special expressions. At first, by differentiating the representation \eqref{PNEW}, and writing $\boldsymbol{\bar{e}}$ instead of $\boldsymbol{ \xi}^{N+1}$, we obtain the $\varrho-$derivative of the pressure function $\mathscr{P}$ as
\begin{align}\label{pderiv}
&  \partial_{\varrho} \mathscr{P}(\varrho, \, \mathbf{q}) = -\frac{\varrho}{q_N} \, \frac{1}{\mathbf{D^{2}}h^*(\mathbf{w^*}) \boldsymbol{\bar{e}}^{} \cdot \boldsymbol{\bar{e}}^{}}\, ,
\end{align}
%
At second, use of \eqref{RHOUNEW} yields
\begin{align}\label{uderiv}
&\partial_{q_N} \mathscr{U}(\varrho, \, \mathbf{q}) =\partial^2_{w^*_{N+1}}h^*(\mathbf{w^*}) - \frac{(\mathbf{D^2}h^*(\mathbf{ w^*}) \mathbf{e}^{N+1} \cdot \boldsymbol{ \bar{e}})^2}{\mathbf{D^{2}}h^*(\mathbf{ w^*})\boldsymbol{ \bar{e}} \cdot \boldsymbol{ \bar{e}}}  \, .
\end{align}
Recall in these formula that $\mathbf{ w^*} = \mathbf{ w^*}(\varrho, \, \mathbf{q})$.
These derivatives possess a sign:
\begin{align}\label{sign}
	\partial_{\varrho} \mathscr{P} > 0 \quad \text{ and } \quad \partial_{q_N} \mathscr{U} > 0 \, .
\end{align}
The quotient $1/(\varrho \, \partial_{\varrho} \mathscr{P})$ possesses the dimension of a compressibility, while $\partial_{q_N} \mathscr{U}/(T^2 \, \varrho)$ has the dimension of a heat capacity.

We can briefly compare these quantities to the compressibility $\beta_{(T,y)}$ and the heat capacity $\calc_{(\upsilon,y)}$ obtained by fixing the composition variable.
Exploiting that $\mathbf{D^2}h(\mathbf{w})$ and $\mathbf{D^{2}}h^*(\mathbf{ w^*})$ are inverse to each other, we can prove for $\mathbf{y_a} = (\boldsymbol{\rho}, \, a)^{\sf T}$ with $a \in \mathbb{R}$ arbitrary that
\begin{align*}
	\varrho = (\mathbf{D^{2}}h(\mathbf{w}))^{\frac{1}{2}} \mathbf{y_a} \cdot (\mathbf{D^2}h^*(\mathbf{ w^*}))^{\frac{1}{2}} \boldsymbol{ \bar{e}} \leq \sqrt{\mathbf{D^{2}}h(\mathbf{w}) \mathbf{y_a} \cdot \mathbf{y_a}} \, \sqrt{\mathbf{D^2}h^*(\mathbf{ w^*}) \boldsymbol{ \bar{e}} \cdot \boldsymbol{\bar{e}}} \, .
\end{align*}
Minimising in $a$ and squaring the result yields
\begin{align*}
	\frac{\varrho^2}{\mathbf{D^{2}}h^*(\mathbf{ w^*}) \boldsymbol{ \bar{e}} \cdot \boldsymbol{ \bar{e}}} \leq \mathbf{D^2} h(\mathbf{w}) (\boldsymbol{\rho},0)^{\sf T}\cdot (\boldsymbol{\rho},0)^{\sf T} - \frac{(\mathbf{D^2} h(\mathbf{w}) (\boldsymbol{\rho},0)^{\sf T}\cdot \mathbf{e}^{N+1})^2 }{\partial^2_{\varrho u}h(\mathbf{w})} \,.
\end{align*}
Now, use of the identities \eqref{relHessiansrho} implies that
\begin{align*}
	\mathbf{D^{2}}h(\mathbf{w}) (\boldsymbol{\rho},0)^{\sf T}\cdot (\boldsymbol{\rho},0)^{\sf T} - \frac{(\mathbf{D^2} h(\mathbf{w}) (\boldsymbol{\rho},0)^{\sf T}\cdot \mathbf{e}^{N+1})^2 }{\partial^2_{\varrho u}h(\mathbf{w})} = \frac{1}{T} \, \mathbf{D^{2}}f(T, \, \boldsymbol{\rho})\boldsymbol{\rho} \cdot\boldsymbol{\rho} \, .
\end{align*}
Thus, invoking \eqref{pderiv}$_1$ with $-1/q_N = T$, we see that
\begin{align*}
	\varrho \, \partial_{\varrho} \mathscr{P} \leq \mathbf{D^{2}}f(T, \, \boldsymbol{\rho})\boldsymbol{\rho} \cdot\boldsymbol{\rho}= \frac{1}{\beta_{(T,y)}} \, .
\end{align*}
Defining $\beta_{(q)}^{-1} := \varrho \, \partial_{\varrho} \mathscr{P}$ as a compressibility as fixed $q$, we obtain  the relationship
\begin{align}\label{bettercompress}
	\beta_{(q)} \geq  \beta_{(T,y)} \, .
\end{align}
%
%
Similarly, we prove in \cite{zbMATH07554819}, equation (71) for the heat capacities that
\begin{align}\label{betterheatcapacity}
	\calc_{(q)} := \frac{1}{T^2 \, \varrho} \,  \partial_{q_N} \mathscr{U} \geq \calc_{(\upsilon,y)} \, .
\end{align}
Note that a more intrinsic notation as $\beta_{(q)}$ is the one used in \eqref{betatrinsic}, with no reference to a particular basis. Clearly $\beta_{(T,\, \mu^t)}$ is the same object as $\beta_{(q)}$.

\subsection{Natural representation with the Gibbs function}

We use \eqref{Gibbsderiv1}, second formula, and \eqref{gibbsyderivpart} to obtain that
\begin{align*}
	\frac{1}{\varrho} = \partial_pg(T, \, p, \, \mathbf{y}), \quad q_i = \frac{d}{dy_i} g (T, \, p, \, \mathbf{y}^\prime) \, \text{ for } \, i = 1, \ldots, N-1
\end{align*}
where we use a particular basis and with $ \mathbf{y}^\prime  = (y_1, \ldots, y_{N-1}, 1-\sum_{i=1}^{N-1} y_i)$. For simplicity, we write from now $ \frac{d}{dy_i} g (T, \, p, \, \mathbf{y}^\prime) = \partial_{y_i} g^\prime$ for $i = 1, \ldots,N-1$, abusing notation.

Regarding $(\varrho, {\bf q})$ as the main variables, we differentiate the latter relations in $\varrho$. 
We obtain that
\begin{align*}
-\frac{1}{\varrho^2} =& \partial^2_p g \, \partial_{\varrho}\mathscr{P} + \mathbf{D^2_{p,\bf y}} g^\prime \cdot \partial_{\varrho} {\bf y}(\varrho, \, \mathbf{q})\, ,\\
0 =& \mathbf{D^{2}_{{\bf y},p}} g^\prime \, \partial_{\varrho}\mathscr{P} + \mathbf{D^{2}_{{\bf y},{\bf y}}} g^\prime \cdot \partial_{\varrho} {\bf y}(\varrho, \, \mathbf{q})\, .
\end{align*}
Here the Gibbs function is evaluated at its natural variables. It follows that
\begin{align*}
	\frac{1}{\varrho^2} = \Big(-\partial^2_{p} g^\prime + (\mathbf{D^{2}_{{\bf y},{\bf y}}} g^\prime)^{-1} \mathbf{D^{2}_{{\bf y},p}} g^\prime \cdot \mathbf{D^2_{{\bf y},p}} g^\prime\Big) \, \partial_{\varrho}\mathscr{P} \, ,
\end{align*}
which we can rewrite (see also \eqref{compressibilityprimaer}) as
\begin{align}\label{kqeasy}
\beta_{(q)} = &\frac{1}{\varrho\,  \partial_{\varrho}\mathscr{P}} = \varrho \, \Big(-\partial^2_{p} g + (\mathbf{D^{2}_{{\bf y},{\bf y}}} g^\prime)^{-1}\, \mathbf{D^{2}_{{\bf y},p}} g^\prime \cdot \mathbf{D^{2}_{{\bf y},p}}  g^\prime\Big)\nonumber\\
= &  -\frac{\partial^2_{p} g}{\partial_p g}  + \frac{(\mathbf{D^{2}_{{\bf y},{\bf y}}} g^\prime)^{-1}\, \mathbf{D^{2}_{{\bf y},p}} g^\prime \cdot \mathbf{D^{2}_{{\bf y},p}}  g^\prime}{\partial_p g} \nonumber\\
= & \beta_{(T,y)} + \frac{ (\mathbf{D^{2}_{{\bf y},{\bf y}}} g^\prime)^{-1}\, \mathbf{D^{2}_{{\bf y},p}} g^\prime \cdot \mathbf{D^{2}_{{\bf y},p}}  g^\prime}{\partial_p g} \, .
\end{align}
Hence, since the latter expression is valid independent of the choice of a basis, we identify
\begin{align}\label{FUND}
\beta_{(T, \mu^t)} - 	\beta_{(T,y)}  = \frac{ (\mathbf{D^{2}_{{\bf y},{\bf y}}} g)^{-1} \,  \mathbf{D^{2}_{{\bf y},p}} g \cdot \mathbf{D^{2}_{{\bf y},p}} g}{\partial_p g} \, ,
\end{align}
as a natural definition of the compressibility of mixing. In fact, due to \eqref{Gibbsderiv1}, equation $2$, in the variables of the Gibbs function \eqref{variables3}, we have
$\partial_\mathbf{y} \varrho/\varrho = -\varrho \, \partial^2_{\bf y,p} g$, and another equivalent representation for \eqref{FUND} is
\begin{align}\label{FUND1}
	\beta_{(T,  \mu^t)} - 	\beta_{(T,y)}  = \partial_pg \,   (\mathbf{D^{2}_{{\bf y},{\bf y}}} g)^{-1} \, \Big(\frac{\partial_{\mathbf{y}} \varrho}{\varrho}\Big) \cdot \Big(\frac{\partial_{\mathbf{y}}  \varrho}{\varrho}\Big)  \, ,
\end{align}
as a measure for how volume changes due to composition. Note that $
\beta_{(T, \boldsymbol{ \mu}^t)} - 	\beta_{(T,y)}$ is another name for $\calk$ defined in \eqref{DEF01}. Here the derivatives with respect to ${\bf y}$ are tangential derivatives.

%
%

\subsection{Explicit formula for ideal mixtures}

For ideal mixtures, the inverse Hessian of the entropy was computed in the Proposition 3.11 of \cite{zbMATH07554819}.
(In these computations we employed rescaled molar masses $M_i \to M_i/R$).  
Recall that $(\mathbf{D^{2}}h(\mathbf{w}))^{-1} = \mathbf{D^2}h^*(\mathbf{ w^*})$, where $\mathbf{ w^*} = \boldsymbol{\nabla}_{\mathbf{w}} h(\mathbf{w})$. In particular
\begin{align}\label{d211}
	\sum_{i,k=1}^N   \partial^2_{w^*_i,w^*_k}h^*
	= \frac{1}{R} \, \sum_{i=1}^N M_i \, \rho_i \,(1-\partial_p g_i \, \varrho)^2 + T \, \varrho^2 \, \sum_{i=1}^N \rho_i \, |\partial^2_pg_i| \, . 
\end{align}
Combination with the identity \eqref{pderiv} yields
\begin{align}\label{kappaqideal}
	\beta_{(T,\mu^t)} = &  \sum_{i=1}^N \rho_i \, |\partial^2_pg_i|+\frac{1}{RT} \, \sum_{i=1}^N M_i \, \rho_i \,\Big(\frac{1}{\varrho}-\partial_p g_i\Big)^2  \nonumber\\
	= & \beta_{(T,y)} +\frac{1}{RT} \,\sum_{i=1}^N M_i \, \rho_i \,\Big(\frac{1}{\varrho}-\partial_p g_i\Big)^2  \, .
\end{align}
With a bit more computational work, we can obtain a similar formula for the heat capacities, with the result $ \calc_{(q)} \geq \calc_{(T,y)}$, with equality only for a pure fluid. Since this point is not at the focus, we spare the computation for the sake of brevity.

\subsection{Incompressibility and constant density}

In the case of an ideal mixture the relation \eqref{kappaqideal} is particularly clear. Interestingly, we can have $\beta_{(q)} = \beta_{(T,y)}$ only in one of the following two cases:
\begin{enumerate}[(i)]
	\item \label{dilute} The mixture reduces to a pure fluid, that is, there is an index $i_0$ such that $\rho_{i_0} = \varrho$ and $\rho_{i} = 0$ for all $i\neq i_0$;
	\item \label{samevol} All species possess the same specific volume $\partial_pg_1(T,p) = \ldots = \partial_pg_N(T,p)$.
\end{enumerate} 
In both cases, the varying composition produces no change in specific volume, and the equation of state \eqref{VOLADD} implies that the pressure is a function of density and temperature, but not of composition.

Using the specific volumes of the species $\partial_p g_i = \calv_i$, a closer look at the nonnegative quantity $$\beta_{(T,\mu^t)}-\beta_{(T,y)} = \frac{1}{RT} \,\sum_{i=1}^N M_i \, \rho_i \,\Big(\frac{1}{\varrho}-\calv_i(T,p)\Big)^2$$ reveals that it precisely measures the volume change of mixing. Indeed, passing to the incompressible limit $\beta_{(T,y)} \rightarrow 0$, which is studied in \cite{zbMATH07824760} and \cite{zbMATH07698266}, implies that $\calv_i(T,\, p)$ must converge to a constant $\bar{\calv}_i$ for each $i$. In the limit, the coefficient $\beta_{(T, \mu^t)}$ does not vanish, but we have
\begin{align}\label{kqincomp}
\beta_{(T, \mu^t)} = \frac{1}{RT} \,\sum_{i=1}^N M_i \, \rho_i \,\Big(\frac{1}{\varrho}-\bar{\calv}_i\Big)^2	\, ,
\end{align}
as a measure of density variations due to composition. Only in the cases \eqref{dilute} and \eqref{samevol} we can expect that $\beta_{(T, \mu^t)}$ is very small too. But these two scenario correspond, in the incompressible limit, precisely to a density $\varrho$ which is nearly constant.

Hence, already in the relatively simple theoretical context of ideal mixtures, we must distinguish between two forms of the incompressible limit:
\begin{enumerate}[(1)]
	\item\label{incomp1} $\beta_{(T,y)} \rightarrow 0$ yielding the incompressibility constraint $\sum_{i=1}^N \bar{\calv}_i \, \rho_i= 1$ which is studied in \cite{zbMATH07824760} and in \cite{zbMATH07330765,zbMATH07452670,zbMATH07698266}. Here a volume/density change due to mixing remains possible;
	\item\label{incomp2} $\beta_{(T, \mu^t)} \rightarrow 0$ yielding the classical incompressibility constraint $\varrho = {\rm Const.}$ with its implication $\divv {\bf v} = 0$ for the velocity field.
\end{enumerate}
Whether \eqref{incomp2} is valid for an incompressible multicomponent fluid depends on the concrete constellation under study. In the next section we shall discuss three illustrations for the relative importance of $\beta_{(T,y)}$ and $\beta_{(T, \mu^t)}-\beta_{(T,y)}$.

The reader interested in yet more thoughts of heuristic nature can consult the appendix, Section \eqref{SoS}, where the coefficient $\beta_{(T, \mu^t)}$ is put in relation to the speed of sound in the fluid.

\section{Numerical examples for three aqueous solutions}\label{EXAMPES}

In this section we consider three binary solutions based on water as the solvent:
\begin{itemize}
	\item Water + salt, seawater;
	\item Water + sucrose, sugar water, syrup;
	\item Water + ethanol.
\end{itemize}
Seawater or ocean water is in fact not a binary mixture. However, in the main theoretical approach, we regard it as a mixture of pure water with a certain {\it salinity}. Salinity quantifies the total amount of dissolved substances -- where ions of sodium and chloride are dominant. 

Also, seawater is not an ideal mixture. At first, it is not possible to rely on data of the pure solute -- neither on data of pure ionic species, nor on data of pure \textquotesingle salt \textquotesingle, which is a granular porous aggregate. However, our strategy is to nevertheless use a binary ideal model, where we fit the data of the solute in order to obtain a good approximation of the true density of seawater {\it in the range of relevant salinity} -- about 3\% mass.

For the water + sucrose case, water can dissolve a much larger quantity of sucrose than of salt. Still, for the same reason as for seawater, we cannot use data of the pure sugar to reliably understand volume effects. We adopt the same strategy as in the first example (to fit an ideal model), in this case with a very good agreement for up to 80\% of sucrose mass.

In the last case of the water + ethanol solution, the mixture exists over the whole range of compositions, but it is far from being volume-additive and thus, it does not exhibit ideal behaviour. Here we discuss both the binary ideal model and a ternary ideal model with one chemical transformation as proposed in \cite{zbMATH07824760}. The result concerning compressibility numbers seems to make sense.

%
%
Considering thus an ideal mixture, we start with a few formula tailored on the binary case. Using \eqref{VOLADD} we have the identity
\begin{align}\label{massideal}
	\sum_{i=1}^N y_i \, \partial_pg_i(T,p) = \frac{1}{\varrho} \, ,
\end{align}
with the mass fractions $y_1,\ldots,y_N$. This allows to re-express
\begin{align}\label{ktyideal}
\beta_{(T,y)} = -\frac{\sum_{i=1}^N y_i \, \partial^2_pg_i(T,p)}{\sum_{i=1}^N	y_i \, \partial_pg_i(T,p)} \, .
\end{align}
For a mixture of two species ${\rm A}_1 = \rm S$ (solute) and ${\rm A}_2 = \rm W$ (water), we denote $y$ the fraction of $\rm S$. Recall that $\partial_pg_i = 1/\hat{\rho}_i$ for $i = 1,2$ with the density $\hat{\rho}_i$ of the constituents as function of temperature and pressure. We introduce $\rho^S := \hat{\rho}_{1}$ and $\rho^W := \hat{\rho}_2$. Then, by means of \eqref{massideal}
\begin{align}\label{densbinaryideal}
	 \varrho = \frac{\rho^W}{1+ y \, (\frac{\rho^W}{\rho^S}-1)} \, ,
\end{align}
Moreover, \eqref{ktyideal} implies that
\begin{align*}
	\beta_{(T,y)} = - \frac{\partial^2_pg^W + y \, (\partial_p^2g^S-\partial_p^2 g^W)}{\frac{1}{\rho^W} + y \, \Big(\frac{1}{\rho^S} - \frac{1}{\rho^W} \Big)}
\end{align*}
With the isothermal compressibility $\beta_{T}^{S,W}$ of the substances, we get
\begin{align}\label{compressidealbinary}
	\beta_{(T,y)} =  \frac{\beta_{T}^{W} + y \, \Big(\frac{\rho^W}{\rho^S}\, \beta_{T}^{S}-\beta_{T}^{W}\Big)}{1+ y \, \Big(\frac{\rho^W}{\rho^S} - 1 \Big)} \, ,
\end{align}
and if the solute is assumed incompressible
\begin{align}\label{compressidealbinaryincomp}
	\beta_{(T,y)} =  \frac{(1-y) \, \beta_{T}^{W}}{1 + y \, \Big(\frac{\rho^W}{\rho^S} - 1 \Big)} \, ,
\end{align}
Similarly, we compute first that
\begin{align*}
\beta_{(q)} - \beta_{(T,y)} =  & \frac{\varrho}{RT} \, \sum_{i=1}^N M_i \, y_i \, \Big(\sum_{k=1}^N y_k \, \partial_pg_k - \partial_pg_i\Big)^2 \\
=  & \frac{1}{RT \, \sum_{k=1}^N y_k \, \partial_pg_k} \, \sum_{i=1}^N M_i \, y_i \, \Big(\sum_{k=1}^N y_k \, \partial_pg_k - \partial_pg_i\Big)^2 \, .
\end{align*}
For the binary mixture, we have
\begin{align*}
\sum_{k=1,2} y_k \, \partial_pg_k - \partial_pg_1 = \frac{y-1}{\rho^S} + \frac{1-y}{\rho^W} \, , \quad \sum_{k=1,2} y_k \, \partial_pg_k - \partial_pg_2 = \frac{y}{\rho^S} - \frac{y}{\rho^W} \, .
\end{align*}
Hence
\begin{align*}
	\sum_{i=1,2} M_i \, y_i \, (\sum_{k=1}^N y_k \, \partial_pg_k - \partial_pg_i)^2 = (M^S \, y \, (1-y)^2 + M^W \, (1-y) \, y^2) \, \Big(\frac{1}{\rho^W}-\frac{1}{\rho^S}\Big)^2 \, .
\end{align*}
Thus, for the difference of the compressibility coefficients we get
\begin{align}\label{compressidealrema}
\beta_{(q)} - \beta_{(T,y)} = \frac{y\, (1-y) \, (M^S+ y\,  (M^W-M^S)) }{R\, T\, \, \rho^W\, (1 + y \, (\rho^W/\rho^S-1))}	\, \Big(1-\frac{\rho^W}{\rho^S}\Big)^2\, .
\end{align} 

\subsection{Seawater}
 
 To obtain data we use the open Python modules {\it seawater 3.3.5} and {\it gsw 3.6.19} (Gibbs Seawater Oceanographic Package) to be obtained from \href{https://pypi.org/project/seawater}{https://pypi.org/project/seawater} and \href{https://pypi.org/project/gsw}{https://pypi.org/project/gsw}. These provide tools to compute several thermodynamic quantities for ocean water from temperature, pressure and salinity input data. In the documentation of the seawater package, we read that \textquotedblleft The package uses the formulas from Unesco’s joint panel on oceanographic tables and standards, UNESCO 1981 and UNESCO 1983 (EOS-80)\textquotedblright{}, to be found in \cite{fofonoff1983algorithms}. Note that this is not the most recent version of the EOS for seawater: \textquotedblleft The EOS-80 library is considered now obsolete; it is provided here for compatibility with old scripts, and to allow a smooth transition to the new TEOS-10.\textquotedblright. We here refer interested readers to \cite{TEOS2010} or the website \href{https://www.teos-10.org}{https://www.teos-10.org} of the project. In particular, the newer equation of state incorporates aspects of thermodynamic consistency by using a Gibbs function as general potential. Several computational routines associated with this approach are implemented in the gsw package. However, the older seawater package offers functions with standard temperature input, which are not yet available in the gsw package. This motivates our choice to use both libraries. 
 
In order to construct the binary ideal model, we at first get the density $\rho^W$ of pure water as a function of temperature and pressure, available in both modules. Next, our strategy is to understand the salinity as an incompressible component, hence $\beta_{T}^S = 0$ (see \eqref{compressidealbinaryincomp}) and $\rho^S$ is a constant, independent on $(T,p)$, in the formulas \eqref{densbinaryideal} and \eqref{compressidealbinaryincomp}. We fit the constant $\rho^S$ in order that the density function \eqref{densbinaryideal} of our ideal model is exact on a standard vector $(T^0,p^0,y^0)$ of reference conditions. Here $T$ is the temperature in degree Celsius, $p$ the pressure in dbar and $y$ the salinity mass fraction.
 
\begin{figure}
\begin{center}	
\includegraphics[width = \textwidth, keepaspectratio = true]{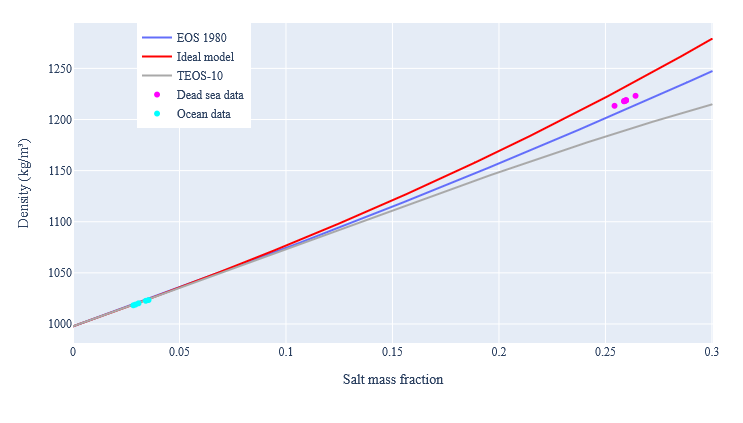}
\captionof{figure}{Density of seawater at reference $T = 25^{\circ} {\rm C}$ and $p =10.1325\, {\rm dbar}$.}
\label{fig:sw_dens}
\end{center}
\end{figure}

To obtain realistic conditions, we use data from {\it Global Seawater Oxygen-18 Levels} which, among other, contains $(T,p,y)$ measurements of ocean conditions done between 1949 and 2009 at different points all over the earth, see \cite{seawaterdata}. After filtering incomplete data, our database still contains $20295$ measurements. Our reference conditions are $3.956$ percent of salinity, a temperature of $24.3^{\,\circ} {\rm C}$ and a pressure of $20.21\,  {\rm dbar}$. After fitting $\rho^S$, we calculate the approximate ideal density using the formula \eqref{densbinaryideal}. We compare the exact density $\varrho^{\rm sw}$ (resp.\ $\varrho^{\rm gsw}$) of the seawater library (resp.\ the Gibbs seawater library) with our calculated density $\varrho^{\rm id}$ according to the ideal model. For the former case, the relative error over the data set is
\begin{align}\label{relativeerr}
{\rm Err}(\varrho) := \frac{\max_{\rm data} |\varrho^{\rm sw} - \varrho^{\rm id}|}{\max_{\rm data} \varrho^{\rm sw} - \min_{\rm data} \varrho^{\rm sw}} = 0.029 \, .
\end{align}
The plot \ref{fig:sw_dens} shows different estimations of the density at reference conditions. For comparison, we also plotted some data for a concentrated solution (dead sea) obtained from \cite{krumgalz1982physico}.
For the ideal model, we hence observe a good fit for the reference ocean waters conditions, and an overestimation of density at larger concentrations. Note that there is no complete agreement between EOS 1980 and TEOS-10 for larger concentrations.
\begin{figure}
	\begin{center}	
		\includegraphics[width = \textwidth, keepaspectratio = true]{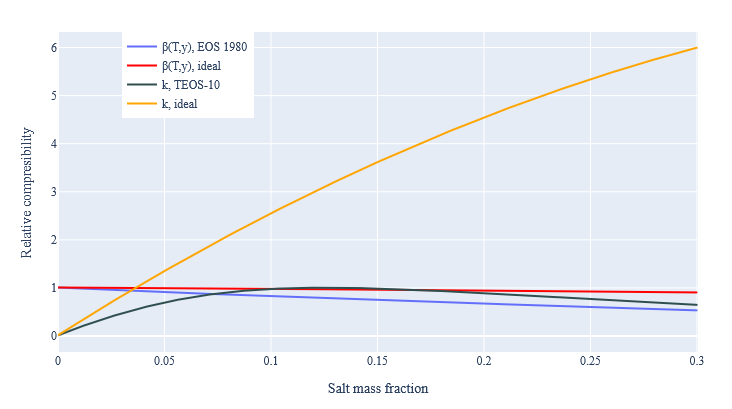}	
		\captionof{figure}{Compressibility of seawater relative to $4.52 \times 10^{-10} \, {\rm  Pa}^{-1}$ at reference $T = 25^{\circ} {\rm C}$ and $p =10.1325\, {\rm dbar}$.}
		\label{fig:sw_comp}
	\end{center}
\end{figure}

As a next step we discuss the compressibility. We evaluate isothermal compressibility according to \eqref{compressidealbinaryincomp} and compare it with a computation using the seawater-module. The latter does not provide the isothermal compressibility directly, but with the available functions: density, speed of sound $U$, heat capacity at constant pressure $\calc_p$ and thermal expansion coefficient $\alpha$, we can compute the exact isothermal compressibility via
\begin{align}\label{heldp}
	\beta^{\rm sw}_{(T,y)} = \frac{1}{\varrho \, U^2} + \frac{\alpha^2 \, T}{\varrho \, \calc_p} \, . 
\end{align}  
We compare the ideal and the exact isothermal compressibility over the data set with the formula \eqref{relativeerr} and we obtain a relatively large error ${\rm Err}(\beta_{(T,y)}) = 0.279$. This is clearly not as good as for the density.

The compressibility of composition $\calk = \beta_{(q)} - \beta_{(T,y)}$ of the ideal model is next evaluated with the formula \eqref{compressidealrema}. In order to obtain a comparison, we use the Gibbs seawater library. It allows to compute the full Gibbs potential as function of $(T,p,y)$ using the formula $g = h - T \, s$, where the specific enthalpy $h$ is given by the function {\it gsw.enthalpy\_t\_exact} and the specific entropy $s$ by {\it gsw.entropy\_from\_t}. Then, we evaluate $\calk$ by means of the formula \eqref{FUND}, where the $(p,y)$-derivatives of $g$ are calculated numerically with a centre-difference scheme. 

The plot \ref{fig:sw_comp} gives the different compressibility numbers relatively to the compressibility of pure water $4.52 \times 10^{-10}\,  {\rm Pa}^{-1}$ as reference value. 
We see that for normal ocean conditions, $\beta_{(q)} - \beta_{(T,y)}$ and $\beta_{(T,y)}$ are of the same small order according to all models. For larger salt contents however, the compositional compressibility $\calk = \beta_{(q)} - \beta_{(T,y)}$ turns to be about six times larger than the isothermal compressibility $\beta_{(T,y)}$ according to the ideal model. However the value obtained from the Gibbs seawater library attains earlier a significantly smaller maximum. Up to which concentration are we allowed to trust the conclusions of the different models? Further investigations will be necessary to better estimate the value of $\calk$.

\subsection{Water and sucrose}

This type of mixture is very important in food industry but also in biology. Our data source for the density of the water+sucrose are:
\begin{itemize}
\item Room temperature, atmospheric pressure in the table on page 5-145 of \cite{CRC}, for mass fractions of solute from $0$ to $80$\%; 
\item Varying temperature, atmospheric pressure in the table C1 of \cite{barbosa2003high}, experimental data;
\item Varying temperature, atmospheric pressure in table 5 of \cite{randall1932ultrasonic}.
\end{itemize}
For the isothermal compressibility, we rely on the values computed in Tables E1-3 of \cite{barbosa2003high}.

Overall our database contains $204$ entries.
\begin{figure}
	\begin{center}	
		\includegraphics[width = \textwidth, keepaspectratio = true]{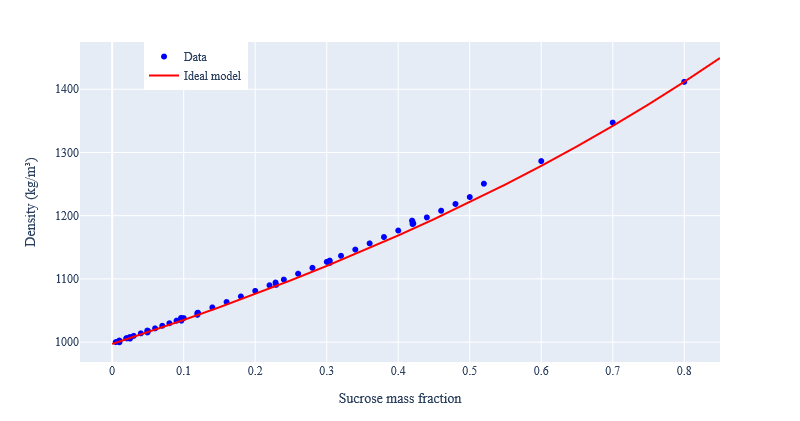}
		
		
		\captionof{figure}{Density of sugar water at reference $T = 25^{\circ} {\rm C}$ and $p =10.1325\, {\rm dbar}$.}
		
		\label{fig:sugw_dens}
	\end{center}

\end{figure}
We fit the density \eqref{densbinaryideal} of our ideal model in order to match the measured density at the reference measurement with $T^0 = 20^{\circ} {\rm C}$, $p^0 = 10.1325 \,  {\rm dbar}$ and $y^0 = 0.8$. The relative error of this density on the (reduced) data set is ${\rm Err}(\varrho) = 0.041$\footnote{We here filter from the data set the pressure-values from \cite{barbosa2003high} exceeding atmospheric pressure of more than $1000$ bars.}. We refer to Figure \ref{fig:sugw_dens} for a plot of the density under reference conditions.

\begin{figure}
	\begin{center}	
		\includegraphics[width = \textwidth, keepaspectratio = true]{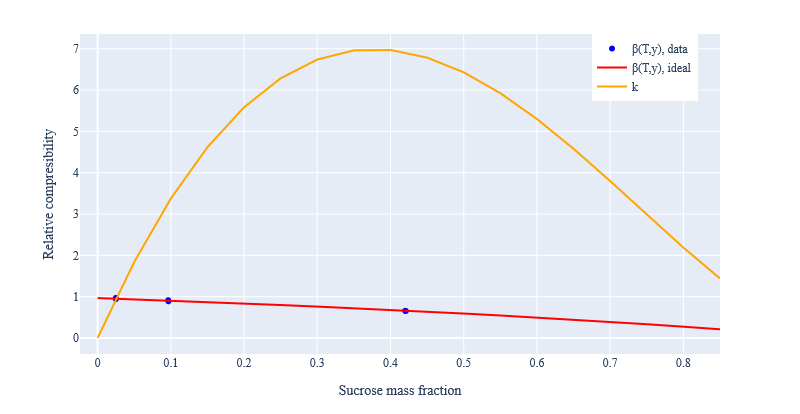}
		
		
		\captionof{figure}{Compressibility of sugarwater relative to $4.52 \times 10^{-10} \, {\rm  Pa}^{-1}$ at reference $T = 25^{\circ} {\rm C}$ and $p =10.1325\, {\rm dbar}$.}
		
		\label{fig:sugw_comp}
	\end{center}
	
\end{figure}

For the isothermal compressibility, the relative error is ${\rm Err}(\beta_{(T,y)}) = 0.072$. We again plot \ref{fig:sugw_comp} the relative compressibilities with respect to the isothermal compressibility of pure water.

We can observe around a mass fraction of sucrose of $0.5$ that the ratio of the compressibility of mixing over the isothermal compressibility attains a maximal value of more than $10$.

\subsection{Water and ethanol}

For this third type of solution, we can rely on data for $\varrho^S$ and $\beta_{T}^S$ for pure ethanol. Our data for the mixture are:
\begin{itemize}
	\item Density data at room temperature, atmospheric pressure in the table on page 5-127, 5-128 of \cite{CRC}, for mass fractions of solute from $0$ to $100$\%; 
	\item Varying temperature and pressure data in the table 2 of \cite{tanaka1977specific}, for mass fractions of solute from $0$ to $100$\%.
\end{itemize}
Our database has $318$ entries. Moreover we have \textquotesingle validated\textquotesingle{} our binary ideal model with respect to these that by using the measurements of excess volumes in \cite{benson1980thermodynamics}.

For the isothermal compressibility, we use the data in table 2 of \cite{tanaka1977specific}, which gives the specific volume for varying pressure, temperature and composition. After fitting a spline of fourth order on the volume data as function of pressure, we can compute its derivative and obtain the isothermal compressibility via $\beta_{(T, \, y)} = -\partial_p\calv/\calv$.

We at first compute the density according to the formula \eqref{densbinaryideal}, where for $\varrho^S = \varrho^S(T,p)$ we use a fit of our data for pure ethanol. The comparison of the ideal density $\varrho^{\rm id}$ to the data yields, even after filtering the data with higher pressure, to a relative error ${\rm Err}(\varrho) = 0.14$. Hence, the ideal model is not really good. Due to the well known effect of volume loss for this mixture, the volume additivity assumption turns critical. 
\begin{figure}
	\begin{center}	
		\includegraphics[width = \textwidth, keepaspectratio = true]{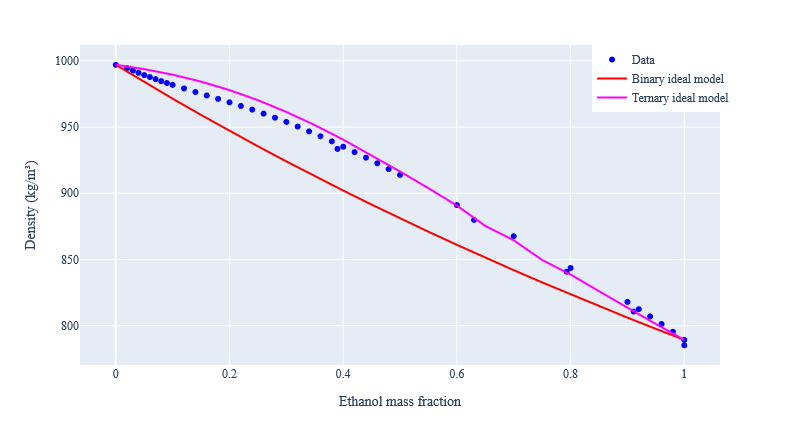}
		
		
		\captionof{figure}{Density of ethanol water near reference $T = 25^{\circ} {\rm C}$ and $p =10.1325\, {\rm dbar}$.}
		
		\label{fig:we_dens}
	\end{center}
\end{figure}
\begin{figure}
	\begin{center}	
		\includegraphics[width = \textwidth, keepaspectratio = true]{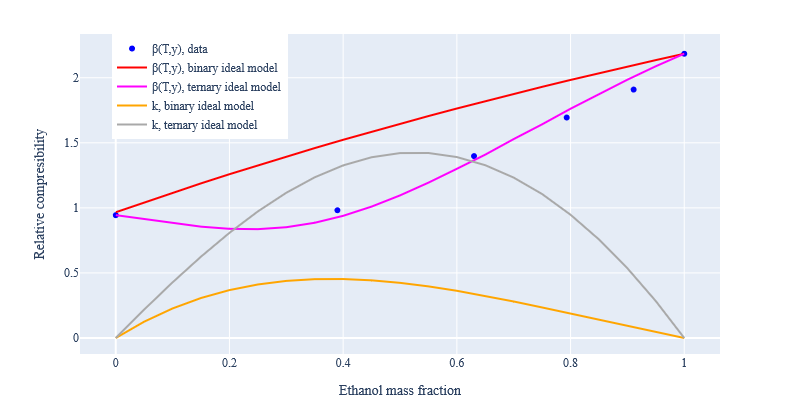}
		
		
		\captionof{figure}{Compressibility of ethanol water relative to $4.52 \times 10^{-10} \, {\rm  Pa}^{-1}$ at reference $T = 25^{\circ} {\rm C}$ and $p =10.1325\, {\rm dbar}$.}
		
		\label{fig:we_comp}
	\end{center}
\end{figure}

For the compressibility, the relative error is even ${\rm Err}(\beta_{(T,y)}) = 0.35$. 

In order to obtain a better estimation of the compressibility coefficients, we switch to a three species ideal model. The procedure is well known for the water+ethanol mixture, see for instance \cite{roux1987association}, Section 3, or Section 5 in \cite{zbMATH07824760}. We have recalled some details of this model in the appendix, Section \ref{ternary}. Results for the density and different compressibilities under standard conditions of temperature and pressure are plotted in the figures \ref{fig:we_dens} and \ref{fig:we_comp}.

Although the density of the ternary ideal model does not completely match the initial slope, it does much better, at least optically, than the binary one.
As to the compressibility, we observe that for the binary ideal model, the compressibility $\calk$ is widely dominated by the isothermal compressibility over the whole range. However, the approximations for the real density and for the isothermal compressibility in this model are rather poor. The ternary model is in the latter respect much better, and it shows that $\calk$ and $\beta_{(T,y)}$ are approximately of the same order. 

Hence for the case of water and ethanol, compressibility effects due on the one hand to pressure and on the other hand to composition look comparable. Approximation of the fluid as a constant density fluid could do well.

\section*{Acknowledgement}

The occasion of the present paper is bidding Prof.~Wolfgang Dreyer farewell, who unfortunately passed away this spring.
I met Wolfgang first in fall 2005 as, fresh graduated from the Humboldt University in Berlin, I started my first position at the Weierstrass Institute, in the group {\it Thermodynamic Modelling and Analysis of Phase-Transitions} of which he was the leader. 
The incompressible limit for fluids was a central topic in our discussions and joined works.
Wolfgang's definition of incompressibility for multicomponent fluids -- as given for instance in \cite{zbMATH06469617}, was thoroughly explained in \cite{zbMATH07452670}, and served as a starting point for several works on analysis, see \cite{zbMATH07330765} or \cite{zbMATH07452670}, \cite{zbMATH07698266}. Basically, incompressibility means zero isothermal compressibility. This definition entails that the mass density (the volume) of an incompressible mixture needs not being constant, which in turn leads to interesting mathematical structures.

Thank you Wolfgang for guiding me through this stimulating journey! 

Last but not least, warm thanks are given to Prof.~Dieter Bothe for his constant scientific support.

\appendix

\section{Differentiating thermodynamic functions of $(\varrho, \, \mathbf{q})$}\label{derivatives} 

At first, by differentiating the representation \eqref{PNEW}, we obtain the $\varrho-$derivative of the pressure function $\mathscr{P}$ as
\begin{align}\label{pderiv2}
	&  \partial_{\varrho} \mathscr{P}(\varrho, \, \mathbf{q}) = -\frac{\varrho}{q_N} \, \frac{1}{\mathbf{D^{2}}h^*(\mathbf{ w^*}) \boldsymbol{ \xi}^{N+1} \cdot \boldsymbol{ \xi}^{N+1}}\, ,\nonumber\\
	& \partial_{q_k} \mathscr{P}(\varrho, \, \mathbf{q}) =- \frac{1}{q_N} \, \Big(\boldsymbol{\nabla}_{\mathbf{ w^*}}h^*(\mathbf{ w^*}) \cdot \boldsymbol{ \xi}^k - \varrho \, \frac{\mathbf{D^{2}}h^*(\mathbf{ w^*})\boldsymbol{ \xi}^{N+1} \cdot \boldsymbol{ \xi}^k}{\mathbf{D^{2}}h^*(\mathbf{ w^*}) \boldsymbol{ \xi}^{N+1} \cdot \boldsymbol{ \xi}^{N+1}} \Big) \quad \text{ for } \quad k =1,\ldots,N-1 \, , \nonumber\\
	& \partial_{q_N} \mathscr{P}(\varrho, \, \mathbf{q}) =- \frac{1}{q_N}\Big(\mathscr{P}+  \partial_{w^*_{N+1}}h^*(\mathbf{ w^*}) - \varrho \, \frac{\mathbf{D^{2}}h^*(\mathbf{ w^*}) \boldsymbol{ \xi}^{N+1} \cdot \mathbf{e}^{N+1}}{\mathbf{D^2}h^*(\mathbf{ w^*}) \boldsymbol{ \xi}^{N+1} \cdot \boldsymbol{ \xi}^{N+1}} \Big)\, .
\end{align}

Using \eqref{RHONEW}, we similarly see for $i = 1,\ldots,N$ and $k = 1,\ldots,N-1$ that
\begin{align}\begin{split}\label{scriptrderiv}
		& \partial_{\varrho} \mathscr{R}_i(\varrho, \, \mathbf{q}) = \frac{\mathbf{D^{2}}h^*(\mathbf{ w^*}) \mathbf{e}^i \cdot \boldsymbol{ \xi}^{N+1}}{\mathbf{D^{2}}h^*(\mathbf{ w^*}) \boldsymbol{ \xi}^{N+1} \cdot \boldsymbol{ \xi}^{N+1}} \, , \\
		& \partial_{q_k} \mathscr{R}_i(\varrho, \, \mathbf{q}) = \mathbf{D^{2}}h^*(\mathbf{ w^*}) \mathbf{e}^i \cdot \boldsymbol{ \xi}^{k} - \frac{\mathbf{D^2}h^*(\mathbf{ w^*}) \mathbf{e}^i \cdot \boldsymbol{ \xi}^{N+1} \, \mathbf{D^{2}}h^*(\mathbf{ w^*}) \boldsymbol{\xi}^{N+1} \cdot \boldsymbol{ \xi}^k}{\mathbf{D^{2}}h^*(\mathbf{ w^*}) \boldsymbol{ \xi}^{N+1} \cdot \boldsymbol{ \xi}^{N+1}}  \, ,\\
		&  \partial_{q_N} \mathscr{R}_i(\varrho, \, \mathbf{q}) = \partial^2_{i,N+1}h^*(\mathbf{ w^*}) - \frac{\mathbf{D^{2}}h^*(\mathbf{ w^*}) \mathbf{e}^i \cdot \boldsymbol{ \xi}^{N+1} \, \mathbf{D^{2}}h^*(\mathbf{ w^*}) \boldsymbol{ \xi}^{N+1} \cdot \mathbf{e}^{N+1}}{\mathbf{D^{2}}h^*(\mathbf{ w^*}) \boldsymbol{\xi}^{N+1} \cdot \boldsymbol{ \xi}^{N+1}}  \, .
	\end{split}
\end{align}

Finally, use of \eqref{RHOUNEW} yields
\begin{align}\label{uderiv}
	& \partial_{\varrho} \mathscr{U}(\varrho, \, \mathbf{q}) = \frac{\mathbf{D^{2}}h^*(\mathbf{ w^*}) \mathbf{e}^{N+1} \cdot \boldsymbol{ \xi}^{N+1}}{\mathbf{D^{2}}h^*(\mathbf{ w^*}) \boldsymbol{ \xi}^{N+1} \cdot \boldsymbol{ \xi}^{N+1}} \, , \nonumber\\
	& \partial_{q_k} \mathscr{U}(\varrho, \, \mathbf{q}) = \mathbf{D^{2}}h^*(\mathbf{ w^*}) \mathbf{e}^{N+1} \cdot \boldsymbol{ \xi}^{k} - \frac{\mathbf{D^{2}}h^*(\mathbf{ w^*}) \mathbf{e}^{N+1} \cdot \boldsymbol{ \xi}^{N+1} \, \mathbf{D^2}h^*(\mathbf{ w^*}) \boldsymbol{ \xi}^{N+1} \cdot \boldsymbol{ \xi}^k}{\mathbf{D^{2}}h^*(\mathbf{ w^*}) \boldsymbol{ \xi}^{N+1} \cdot \boldsymbol{ \xi}^{N+1}} \, ,\nonumber\\
	&\partial_{q_N} \mathscr{U}(\varrho, \, \mathbf{q}) =\partial^2_{N+1}h^*(\mathbf{ w^*}) - \frac{(\mathbf{D^{2}}h^*(\mathbf{ w^*}) \mathbf{e}^{N+1} \cdot \boldsymbol{ \xi}^{N+1})^2}{\mathbf{D^{2}}h^*(\mathbf{ w^*}) \boldsymbol{ \xi}^{N+1} \cdot \boldsymbol{ \xi}^{N+1}}  \, .
\end{align}
We can verify the following the relationships
\begin{align*}
	\partial_{q_N} \mathscr{M} = -\partial_{\varrho} \mathscr{U} \, \quad 
	\partial_{\varrho} \mathscr{M} = T \, \varrho \, \partial_{\varrho} \mathscr{P}\\
	\partial_{q_N} \mathscr{P} = T \, (\mathscr{P} +\mathscr{U} - \varrho \, \partial_{\varrho} \mathscr{U}) \, .
\end{align*}

\section{Wave equation for a multicomponent fluid}\label{SoS}

In a fluid near equilibrium (zero velocity and constant thermodynamic state), compression waves might propagate. As is well known,  the speed of compression waves is underestimated by the isothermal compressibility number, as the temperature gradient can align with the density gradient to accelerate the wave. The speed of sound is thus $1/(\varrho \, \beta_{s})$ with the isentropic (adiabatic) compressibility $\beta_{s}$.

This regime is described for instance in Section 10.4 \cite{muller2013grundzuge} for a single component fluid in one space dimension. Here we shall briefly consider in the same spirit the multicomponent case, where in addition mass diffusion gradients can also align with the pressure gradient.

We start from the partial differential equations of multicomponent fluid dynamics where diffusion fluxes, viscosity and external forces are overall neglected. The equations are posed in $\mathbb{R}^3 \times (0,+\infty)$, the position variable is denoted by $\calx = (\calx_1,\calx_2,\calx_3)$, the time is $t$. Thus
\begin{align}
	&	\label{mass} \partial_t\rho_i +\divv(\rho_i \,  {\bf v}) = 0\\
	&	\label{energy} \partial_t\varrho u +\divv(\varrho u \,  {\bf v}) = - p \, \divv {\bf v}\\
	&	\label{momentum} \partial_t\varrho {\bf v} +\divv(\varrho \, {\bf v} \otimes  {\bf v}) + \boldsymbol{\nabla_\calx} p = 0 
\end{align}
with the velocity field ${\bf v} = (v_1,v_2,v_3)$. We look for solutions in the form of small perturbations of a constant state solution: $\rho_i =\rho_i^0 + \bar{\rho}_i$, $T = T^0 + \bar{T}$, with a velocity field ${\bf v} = (\bar{v}_1,\bar{v}_2,\bar{v}_3)$ which is a small perturbation of zero. Notice that not only $T$  and $\boldsymbol{\rho}$ are small perturbation of constant values, but also other thermodynamic variables $\boldsymbol{ \mu}$, $\boldsymbol{y}$, etc.

After dropping from the equations \eqref{mass}, \eqref{energy}, \eqref{momentum} all products of perturbations which are second order small, we deduce in well known manner from \eqref{mass} and \eqref{momentum} the equation $\partial^2_t\varrho - \Delta p = 0$. We use the vectors $\mathbf{w} = (\boldsymbol{\rho}, \, \varrho u)$ to rewrite \eqref{mass}, \eqref{energy} as
\begin{align*}
	\dot{\mathbf{w}} = - (\boldsymbol{\rho}, \, \varrho u+p)^{\sf T} \, \divv {\bf v} \, ,
	%
\end{align*}
and since ${\bf w} = \boldsymbol{\partial}_{\mathbf{w^*}} h^*(\mathbf{w^*})$ also
\begin{align*}
	\dot{\mathbf{ w^*}} = - \mathbf{D^{2}}h(\mathbf{w}) \, (\boldsymbol{\rho}, \, \varrho u+p)^{\sf T} \, \divv {\bf v} \, . 
\end{align*}
Due to the continuity equation, we have $-\varrho \, \divv {\bf v} = \dot{\varrho}$ and hence
\begin{align*}
	\dot{\mathbf{ w^*}} = \frac{1}{\varrho} \, \mathbf{D^{2}}h(\mathbf{w}) \, (\boldsymbol{\rho}, \, \varrho u+p)^{\sf T}\, \dot{\varrho} \, .
\end{align*}
Using that the perturbations vanish for large time and integrating the latter yields
\begin{align*}
	\mathbf{\bar{w}^*} = \frac{1}{\varrho^0} \,  \mathbf{D^{2}}h(\mathbf{w}^0) \, (\boldsymbol{\rho}^0, \, \varrho^0 u^0+p^0)^{\sf T} \, \bar{\varrho} \, .
\end{align*}
Applying the spacial gradient yields
\begin{align}\label{allalign}
	\boldsymbol{\nabla_{\calx}} \mathbf{w^*} = \frac{1}{\varrho^0} \,  \mathbf{D^{2}}h(\mathbf{w}^0) \, (\boldsymbol{\rho}^0, \, \varrho^0 u^0+p^0)^{\sf T} \, \boldsymbol{\nabla_{\calx}}\bar{\varrho} \, ,
\end{align}
and we see that the gradients of all thermodynamic variables align to the density gradient. The identity $p = T \, h^*(\mathbf{ w^*}) = - h^*(\mathbf{ w^*}) / w_{N+1}^*$ is next differentiated, and we get
\begin{align*}
	\boldsymbol{\nabla_{\calx}} p = \frac{T^0}{\varrho^0} \, \mathbf{D^2}h(\mathbf{w}^0) (\boldsymbol{\rho}^0, \, \varrho^0 u^0+p^0)^{\sf T}\cdot (\boldsymbol{\rho}^0, \, \varrho^0 u^0+p^0)^{\sf T}\, \boldsymbol{\nabla_{\calx}} \varrho \, .
\end{align*}
Hence with 
\begin{align}\label{speedofsound}
	c^2 = \frac{T}{\varrho} \, \mathbf{D^2}h(\mathbf{w}) (\boldsymbol{\rho}, \, \varrho u+p)^{\sf T}\cdot (\boldsymbol{\rho}, \, \varrho u+p)^{\sf T}
\end{align}
we have identified the speed of compression waves in the fluid. 

\textquotedblleft Compression waves \textquotedblright{} appear in this argument \eqref{allalign} to be be accelerated by mass and heat diffusion processes. To show this, we use the function $\mathscr{P}$ to compute that
\begin{align*}
	\boldsymbol{\nabla_{\calx}} p = &  \mathscr{P}_{\varrho}(\varrho, \, \mathbf{q}) \, 
	\boldsymbol{\nabla_{\calx}}  \varrho +	\mathscr{P}_{\mathbf{q}}(\varrho, \, \mathbf{q}) \cdot 
	\boldsymbol{\nabla_{\calx}}  \mathbf{q} \\
	= & \frac{1}{\varrho\, \beta_{(q)}} \, 
	\boldsymbol{\nabla_{\calx}} \varrho + 	\mathscr{P}_{\mathbf{q}}(\varrho, \, \mathbf{q}) \cdot 
	\boldsymbol{\nabla_{\calx}} \mathbf{q} \, .
\end{align*}
We can interpret $
\boldsymbol{\nabla_{\calx}}  q_1, \ldots,
\boldsymbol{\nabla_{\calx}}  q_N$ as diffusion driving forces. Mainly $
\boldsymbol{\nabla_{\calx}}  q_1, \ldots,
\boldsymbol{\nabla_{\calx}}  q_{N-1}$ are driving the mass diffusion while $
\boldsymbol{\nabla_{\calx}}  q_N$ is responsible for the heat diffusion -- provided that cross effects named after Soret and Dufour are neglected. In this case, the Maxwell-Stefan equations relate the mass diffusion fluxes $\mathbf{j}_1, \ldots, \mathbf{j}_N$ and the driving forces via
\begin{equation}\label{MS-mass-based}
	- \sum_{k=1}^{N} f_{ik} \, (y_k \, \mathbf{j}_i - y_i \, \mathbf{j}_k )   = \rho_i \, \big( 
	\boldsymbol{\nabla_{\calx}}  \frac{\mu_i}{RT} - \sum_{k=1}^{N} y_k \, 
	\boldsymbol{\nabla_{\calx}}  \frac{\mu_k}{RT} \big) \quad \mbox{ for } i=1,\ldots ,N \, ,
\end{equation}
in which $\{f_{ik}\}$ is a matrix of phenomenological coefficients describing the friction between the species (reciprocal diffusivities), and $R$ is the gas-constant. In tensorial form 
\begin{align}\label{msfuerdich}
	\mathcal{B} \, \mathbf{j} = - \mathcal{R} \, \mathcal{P} \,
	\boldsymbol{\nabla_{\calx}}   \frac{\boldsymbol{ \mu}}{T}  \, \quad \text{ 
		with } \quad  \mathcal{B}_{ik} = \begin{cases}
		-f_{ik} \, y_i & \text{ for } i \neq k\\
		\sum_{j\neq i} f_{ij} \, y_j & \text{ for } i = k \, ,
	\end{cases}
\end{align}
with $\mathcal{R} := \text{diag}(\rho_1,\ldots,\rho_N)$ and $\mathcal{P} := \boldsymbol{\rm Id} - {\bf 1} \otimes \mathbf{y}$, with ${\bf 1} = 1^N$.
Using the properties \eqref{etaspecia} of the basis $\{\eta^{\ell}\}$ and the properties of $\mathcal{B}$ we have $-\boldsymbol{\nabla}_{\calx} q_\ell = \boldsymbol{\eta}^{\ell} \cdot \mathcal{R}^{-1} \, \mathcal{B} \, \mathbf{j}$ for $\ell = 1,\ldots,N-1$. By means of \eqref{pderiv2} we then see that
\begin{align}\label{pressurediff}
	\sum_{\ell = 1}^{N-1}	\mathscr{P}_{q_\ell}(\varrho, \, \mathbf{q}) \, 
	\boldsymbol{\nabla_{\calx}}  q_{\ell} = -\frac{T\, \varrho}{\mathbf{D^{2}}h^* \boldsymbol{ \xi}^{N+1} \cdot \boldsymbol{ \xi}^{N+1}} \, \mathcal{B}^{\sf T} \, \mathcal{R}^{-1} \, \mathbf{D^{2}}h^* \boldsymbol{ \xi}^{N+1}  \cdot \mathbf{j} \, .
\end{align}
Similarly, if the heat diffusion flux is given by a Fourier law $\mathbf{j}^h = -\kappa \, 
\boldsymbol{\nabla_{\calx}}  T$, then use of \eqref{pderiv2} yields
\begin{align}\label{pressureheatdiff}
	\mathscr{P}_{q_N}(\varrho, \, \mathbf{q}) \, 
	\boldsymbol{\nabla_{\calx}}  q_{N} & = -\frac{\mathbf{j}^h}{\kappa \, T^2} \, \mathscr{P}_{q_N}(\varrho, \, \mathbf{q})\nonumber \\
	& = -\frac{\varrho u + p - \varrho \, \mathbf{D^{2}}h^* \boldsymbol{ \xi}^{N+1}\cdot \mathbf{e}^{N+1}/\mathbf{D^{2}}h^* \boldsymbol{ \xi}^{N+1}\cdot \boldsymbol{ \xi}^{N+1}}{\kappa \, T} \, \mathbf{j}^h \, . 
\end{align}
To resume, a general decomposition of the pressure gradient yields a compressive contribution and a diffusive enhancement:
\begin{align*}
	\boldsymbol{\nabla_{\calx}} p = &   \frac{1}{\varrho\, \beta_{(q)}} \, 
	\boldsymbol{\nabla_{\calx}} \varrho   -\frac{T\, \varrho \, \mathcal{B}^{\sf T} \, \mathcal{R}^{-1} \, \mathbf{D^{2}}h^* \boldsymbol{ \bar{e}} }{\mathbf{D^{2}}h^* \boldsymbol{ \bar{e}} \cdot \boldsymbol{ \bar{e}}} \cdot \mathbf{j}\\
	& -\frac{\varrho u + p - \varrho \, \mathbf{D^{2}}h^* \boldsymbol{ \bar{e}}\cdot \mathbf{e}^{N+1}/\mathbf{D^{2}}h^* \boldsymbol{ \bar{e}}\cdot \boldsymbol{ \bar{e}}}{\kappa \, T} \, \mathbf{j}^h \, .
\end{align*}

The compressive part is expressed by the total compressibility coefficient $\beta_{(q)}$ rather than by the more usual isothermal compressibility.

\section{Water+ethanol mixture as ternary ideal system}\label{ternary}

We follow closely the section 5 of \cite{zbMATH07824760}. See also \cite{roux1987association}, section 3 for a variant of the same model. Through a chemical transformation $\gamma_1 \, {\rm A}_1 + \gamma_2 \, {\rm A}_2 \rightarrow \gamma_3 \, {\rm A}_3$ the two species combine to a third one. Here the numbers $\gamma_i$ are positive integers. The transformation conserve mass, hence with $a_i = \gamma_i/\gamma_3$ for $i = 1,2$, we must have $M_3 = a_1 \, M_1 + a_2 \, M_2$. However, since the volume needs not to be conserved, the model can explain deviation from volume additivity of the binary ideal model.

The system is assumed in chemical equilibrium, which is satisfied if $\gamma_1M_1 \, \mu_1+\gamma_2M_2 \, \mu_2 = \gamma_3M_3 \, \mu_3$. More simply, with $\theta = a_2M_2/(a_1 \, M_1 + a_2 \, M_2)$
\begin{align*}
	\mu_3 = (1-\theta)\, \mu_1 + \theta \, \mu_2\, .
\end{align*}
Assuming now the ideal form \eqref{muiideal} of the chemical potentials we can solve the latter equation for the mole fraction $x_3$:
\begin{align*}
	x_3 = x_1^{a_1} \, x_2^{a_2} \, \exp\Big(\frac{M_3}{RT} \, \Big((1-\theta) \, g_1(T,p) + \theta \, g_2(T,p) - g_3(T,p)\Big)\Big) \, .
\end{align*}
The function under the exponential represents the deviation from volume additivity. Since we need to fit both the density and the compressibility, we make a quadratic Ansatz
\begin{align*}
 \frac{M_3}{RT} \, \Big((1-\theta) \, g_1(T,p) + \theta \, g_2(T,p) - g_3(T,p)\Big) =:	& \tilde{g}(T,p) \\
&  = (p-p_0) \, (D(T) + D^\prime(T) \, (p-p_0)/2) \, ,
\end{align*}
with two functions $D(T), \, D^\prime(T)$ to be determined. Hence, we will have
\begin{align*}
g_3(T,p) = (1-\theta) \, g_1(T,p) + \theta \, g_2(T,p) - \frac{RT}{M_3} \, (p-p_0) \, (D(T) +D^\prime(T) \, (p-p_0)/2) \, ,
\end{align*}
and have complete data for an ideal ternary model. Using the variables $y = $ mass fraction of ${\rm A}_2$, $T$ and $p$, we can parametrise the complete state. Indeed, we obtain the three mole fractions $x = x(y,T,p)$ by solving for $x$ the three equations
\begin{align}\label{shoud}
x_3 = x_1^{a_1} \, x_2^{a_2} \, e^{\tilde{g}(T,p)}, \quad M_2\, x_2 = y \, \sum_{i= 1}^3M_i \, x_i,\quad x_3 = 1-x_1-x_2 \, .
\end{align}
Straightforward calculations and \eqref{VOLADD} yield
\begin{align}\label{volfit}
\frac{1}{\varrho} = & (1-y)\, \partial_pg_1(T,p) + y \, \partial_pg_2(T,p)\nonumber \\
&+ y_3(y,T,p) \, (\theta \,\partial_p(g_2-g_1)(T,p) - \frac{RT}{M_3} \, (D(T)+D^\prime(T)\, (p-p_0))) \, .
\end{align}
Here $y_3(y,T,p)$ is obtained via solution of \eqref{shoud}. Having measurement $V_0(T)$ for the volume of the binary mixture at some reference point $(y_0,p_0)$, we get for $D$ the equation
\begin{align}\label{volfit2}
	V_0(T) = & (1-y_0)\, \partial_pg_1(T,p_0) + y_0 \, \partial_pg_2(T,p_0) \nonumber\\
	& + y_3(y_0,T,p_0) \, \Big(\theta \,\partial_p(g_2-g_1)(T,p_0) - \frac{RT}{M_3} \, D(T)\Big) \, .
\end{align}
Moreover, using differentiation in $p$ of \eqref{volfit}, we can compute the relation
\begin{align}\label{compfit}
\frac{\beta_{(T,y)}}{\rho} =& (1-y)\, \beta_T^1\,  \partial_pg_1 + y \,\beta_T^2\,  \partial_pg_2 + y_3 \, \Big(\theta \,(\beta_T^2\partial_pg_2- \beta_T^1\, \partial_pg_1) - \frac{RT}{M_3} \, D^\prime(T)\Big)\nonumber\\
&	+  \partial_p y_3 \, \Big(\theta \,\partial_p(g_2-g_1) - \frac{RT}{M_3} \, (D+D^\prime\, (p-p_0))\Big)
	 \, . 	
\end{align}
with the compressibilities $\beta_T^i$ of the pure species. Restricting to a reference state $(y_0,p_0)$ for which we possess a measurment of the true compressibility of the binary mixture, we determine $D^\prime(T)$. Hence a ternary model is finally fitted.

\end{document}